\documentclass{article}
\usepackage{graphicx} 
\usepackage{subfigure}
\usepackage{bm}
\usepackage{amsmath}
\usepackage{amssymb}
\usepackage{xcolor}
\usepackage{amsthm} 
\usepackage{mathtools}
\usepackage{authblk}
\usepackage{hyperref}
\usepackage{natbib}
\usepackage{orcidlink}

\title{{Propagation of elastic waves in a flexomagnetic solid}\footnote{Notice: This manuscript has been coauthored by UT-Battelle, LLC, under Contract No. DE-AC0500OR22725 with the U.S. Department of Energy. The United States Government retains and the publisher, by accepting the article for publication, acknowledges that the United States Government retains a non-exclusive, paid-up, irrevocable, world-wide license to publish or reproduce the published form of this manuscript, or allow others to do so, for the United States Government purposes. The Department of Energy will provide public access to these results of federally sponsored research in accordance with the DOE Public Access Plan (\href{http://energy.gov/downloads/doe-public-access-plan}{http://energy.gov/downloads/doe-public-access-plan}).}}
\author[1]{Swarnava Ghosh\orcidlink{0000-0003-3800-5264}}
\affil[1]{National Center for Computational Sciences, Oak Ridge National Laboratory, TN 37830 \\ Email: ghoshs@ornl.gov}
\begin{document}

\maketitle

\begin{abstract}

Flexomagnetism is the coupling between magnetism and strain gradients and is a technologically relevant phenomenon. We present a theory of elastic wave propagation in a linear elastic flexomagnetic material with microstructure and strain gradient elastic interactions. The expressions of frequency, phase velocity, and group velocity of longitudinal and transverse waves are derived and are shown to depend on the flexomagnetic coefficient and microstructure. We also show that the effect of flexomagnetism and microstructure can lead to some interesting phenomena in wave propagation, which are not observed in classical linear elasticity theory of waves. Specifically, in contrast to classical linear elastic materials, where wave propagation is non-dispersive, flexomagnetic materials with microstructure can exhibit both normal and abnormal dispersion. It is also noteworthy that, in flexomagnetic materials with gradient elasticity, the phase velocities of transverse waves can exceed those of longitudinal waves, which is atypical in classical elasticity. Furthermore, waves can also attenuate for a certain range of wavenumbers that depend on the flexomagnetic coefficient and microstructural parameters. Finally, we explore the possibility of waves exhibiting zero group velocity modes, where waves are non-propagating but have strong local energy confinement, negative group velocity modes, where the wave packet moves in the opposite direction to that of wave propagation, and the phenomenon of wave freezing, where a propagating wave stops in space without diffusing or spreading.

\end{abstract}
\section{Introduction}
Flexomagnetism is the property by which strain gradients in materials induce magnetism, without an external magnetic field, or a temporally varying electric field, or electric currents (\cite{eliseev2009,Eliseev:PRB,sky2025cosserat}). It is a technologically relevant magnetomechanical phenomenon and can potentially revolutionize the development of sensors, actuators, energy harvesters, and data storage by manipulating magnetic fields in materials with strain gradients (\cite{tang2025flexomagnetism}). Flexomagnetic effect diminishes with the increase in size, and therefore it's influence on the magneto-mechanical response is significant in nanoscale materials.  

The analogous phenomenon of flexoelectricity, i.e., coupling between strain gradient and electric polarization, is a universal property of all dielectrics (\cite{Yudin2013Review,Tripathy2021Review,Codony2021Review,Nguyen2013Review,Wang2016Review}), and has been the subject of investigations in recent years (\cite{Sharma2007,Abdollahi:2014,Codony2021Review}). In particular, flexoelectricity gives rise to interesting observations in thin films and two-dimensional materials (\cite{ghosh2026generalization,kumar2025,codony2021,Kothari2018,Kothari2019}), soft matter (\cite{Deng2014,Grasinger2021,codony:Finite}), biological materials (\cite{Deng2019,Gao2008,Witt2023}), and liquid crystals (\cite{Rahmati2025}).

Some important distinctions between the two flexo effects are that flexomagnetism requires long-range magnetic order, whereas long-range electrical ordering is not required for materials to exhibit flexoelectricity (\cite{sky2025cosserat}). Another distinction is the atomic scale origins of flexoelectricity and flexomagnetism. In an electrically polarized material, the electric dipoles originate from spatially separated charges (\cite{Sharma2007}). Still, in a magnetized material, magnetic moments do not arise from the spatial separation of magnetic poles (\cite{coey2010magnetism}). Another key difference between the two flexo effects is the type of symmetry breaking associated with strain gradients. Flexoelectricity arises from the breaking of inversion symmetry, whereas flexomagnetism originates from the breaking of time-reversal symmetry, thereby producing a magnetic response (\cite{tang2025flexomagnetism}).

Flexomagnetism can arise from several distinct nanoscale mechanisms. In particular, nonuniform strain fields can directly modify atomic magnetic moments, resulting in a net change in magnetization. Strain gradients can also alter magnetic anisotropy and magnetic exchange interactions, and may induce asymmetric exchange couplings, which together lead to measurable variations in the magnetization of a material. We review some of the developments in flexomagnetism below.

\cite{bobylev1979} observed coupled electromagnetic and elastic effects due to strain gradients in liquid crystals. \cite{eliseev2009} developed a theory of flexomagnetism and flexoelectricity of ferroic systems based on the Landau–Ginzburg–Devonshire framework. They showed that flexoeffects strongly influence atomic displacements, symmetry changes, and phase-transition temperatures. In a follow-up work, \cite{Eliseev:PRB} showed that the flexomagnetic effect gives rise to new linear flexomagnetoelectric coupling in ferroics, where the polarization and (anti)magnetization vectors are spatially inhomogeneous. The flexomagnetoelectric coupling between the polarization and magnetization strongly influences the dielectric susceptibility and ME tunability of multiferroics. Using first-principles calculations, \cite{lukashev2010flexomagnetic} showed that strain gradients lead to a net magnetization in antiferromagnetic Mn$_3$GaN. Interestingly, in BiFeO$_3$, strain gradients alter magnetocrystalline anisotropy, giving rise to the flexomagnetic effect \cite{Lee2017}.

\cite{hertel2013curvature} showed that curvature, as seen in nanotubes and flexible membranes, leads to surprising effects when the radius of curvature is of the order of intrinsic length scales, such as domain wall width or the magnon wavelength. Simulations show that curved ferromagnetic thin films display magnetochiral properties similar to the Dzyaloshinskii-Moriya interaction (DMI) (\cite{dzyaloshinsky1958thermodynamic,moriya1960anisotropic}), suggesting that controlled bending of ferromagnetic membranes can provide a reversible and universal method to tune their magnetic properties. Curvature-driven magnetochiral effect in soft ferromagnetic materials has also been observed experimentally (\cite{volkov2019}). Furthermore, their work also quantified the effective DMI constant, whose value can surprisingly be as high as interfacial DMI constants in thin films. Further building on the research in curved magnetism, \cite{edstrom2022curved} showed that CrI$_3$, a typical two-dimensional Ising ferromagnetic insulator with vertical easy axis, is an ideal material for realizing flexomagnetic phenomena. They used a bent ring structure for first-principles calculations and determined the curvature-dependent energy of several magnetic states. This analysis was further extended to an antiferromagnetic bilayer of CrI$_3$ (\cite{qiao2024curvature}). Curvature in MnSe$_2$ can introduce DMI that leads to chiral spin textures such as magnetic skyrmions (\cite{ga2022}). Beyond two-dimensional materials, flexomagnetic effects are also observed in thin antiferromagnetic GbPtSb wrinkled membranes, which show spontaneous magnetization (\cite{du2021epitaxy}), and a wrinkled thin film of paramagnetic CoCrFeMnNi high entropy alloy showing an increase in magnetization (\cite{ling2023}).

Remarkably, flexomagnetism can influence magnetic ordering temperatures in materials. For example, in a heterostructure of ferromagnetic LaSrMnO$_3$ and piezoelectric PbZrTiO$_3$, lattice distortion due to crystal mismatch generates strain gradients that alter the Curie temperature (\cite{spurgeon2014}). In antiferromagnetic Cr$_2$O$_3$ thin films, strain gradients generate a vertically graded Néel temperature profile \cite{makushko2022}. In a thin film of ferromagnetic SrRuO$_3$, epitaxially grown on a substrate, shear strain gradient enhances magnetization and alters the Curie temperature \cite{peng2024}.

These advances have also prompted the development of continuum field theories of flexomagnetism. \cite{sidhardh2018} developed a constitutive model of flexomagnetic effect in nano-beams for both converse and direct effects. \cite{malikan2020geometrically} presented an analysis of geometrically nonlinear vibrations of piezo-flexomagnetic nanotubes, and \cite{zhang2022size} analyzed the free vibration and buckling of flexomagnetic nanobeams considering both direct and converse effects. \cite{malikan2020instabilities} investigated instabilities and post-buckling behavior of piezomagnetic nanostructures exhibiting flexoelectricity, \cite{malikan2021effect} developed a theory of geometrically nonlinear bending of the Euler-Bernoulli ferromagnetic nanobeams accounting for surface effects, and \cite{malikan2022flexomagneticity} investigated flexomagnetism in a shear deformable piezomagnetic nanostructure. Recently, a geometrically nonlinear Cosserat micropolar finite continuum theory of flexomagnetism has been developed by \cite{sky2025cosserat}, thereby extending the continuum theory of flexomagnetism to finite deformation regimes. 

The propagation of waves in materials underpins a wide range of engineering and scientific applications such as nondestructive evaluation of materials and structures for stress and damage(\cite{cawley1996use,rose2014ultrasonic}), vibration isolation and wave guiding in phononic crystals and metamaterials (\cite{kushwaha1993acoustic,hussein2014dynamics}), seismic imaging and geophysical exploration (\cite{aki2002quantitative}), sensing using waves in micro- and nano-electro-mechanical systems (\cite{campbell1998surface,morgan2010surface}).  

Also of interest is the effect of microstructure, which can impart fascinating features to materials (\cite{bhattacharya2003microstructure}). The extension of linear elasticity to include microstructure was first formulated by \cite{Mindlin1964}. Subsequently, \cite{eringen1966linear} developed a theory of micropolar elasticity, and \cite{toupin1964theories,koiter1969couple} formulated the couple stress theory. \cite{mindlin1968first} further extended his theory to include strain gradient effects in the energy, which can arise due to non-local interactions. In these theories, the effect of microstructure and non-local interactions is effectively accounted for by length parameters. \cite{Papargyri2009} developed a unified theory for waves traveling in a linear elastic medium with microstructure and non-local effects. Notably, their theory predicts wave-propagation phenomena absent in classical elasticity, including nonlinear dispersion relations and wavenumber ranges over which the waves become nonphysical. Another interesting phenomenon related to waves observed in materials with microstructure is that of negative group velocity modes, where the wave packet moves in the opposite direction to that of wave propagation (\cite{meitzler1965,dolling2006simultaneous,martinez1984negative,ye2013negative}), zero group velocity modes, where waves are not propagating but energy is confined locally (\cite{prada2008local,prada2009influence,kausel2012number}), and wave freezing, where a propagating wave stops in space without diffusing or spreading (\cite{figotin2006frozen,Fishman2024,Deshmukh2024}).
 
Elastic wave propagation in flexoelectric materials has been extensively investigated, and studies indicate that flexoelectricity gives rise to effects that are not typically observed in waves propagating in classical elastic materials. For example, in a non-piezoelectric material, transverse shear surface acoustic waves can propagate due to flexoelectricity \cite{Eliseev2017}. \cite{Hu:2018} developed a theoretical framework for the propagation of longitudinal elastic waves in a semi-infinite flexoelectric microstructured solid. Their analysis demonstrates that, unlike classical elasticity—which predicts nondispersive wave propagation—the presence of flexoelectricity leads to dispersive waves, with both phase and group velocities depending on the wavenumber. \cite{Giannakopoulos2022} developed a theoretical framework that includes magnetic effects and demonstrated that the electric field is linearly related to the dilatational components of particle acceleration, while the magnetic flux and magnetic field are proportional to the shear components of particle velocity. These relations provide a basis for studying seismo-electromagnetic phenomena. Following up on this, \cite{Giannakopoulos2024} presented an analysis of the dispersion and attenuation of waves in a flexoelectric medium with microstructure and gradient elasticity. They showed that both longitudinal and shear waves are dispersive, and group and particle velocities depend on the flexoelectric coefficient as well as microstructural and gradient elastic length scales. Furthermore, at a certain range of wave numbers, the waves can also attenuate.  \cite{Seema2025} analyzed the elastic wave propagation across the interface of a visco-piezo composite made from lanthanum niobate and aluminum nitride, loosely bonded to a piezoelectric substrate exhibiting flexoelectricity. They show that aluminum nitride shows higher phase velocities, while lanthanum niobate influences the attenuation of these waves. Overall, the flexoelectric effect has a significant influence on the wave propagation in both materials, impacting both velocities and attenuation.

We also mention some contemporaneous investigations on waves in materials with flexomagnetic effect. \cite{hrytsyna2023love} analyzed the effects of flexomagnetism, flexoelectricity, and microinertia on the propagation of Love waves in layered materials. Their results show that dispersion curves depend on the layered structure's composition, layer thickness, and the relative values of the flexo-coefficients. \cite{jiao2024dispersion} studied the dispersion and reflection of coupled waves in a piezo-flexo-magnetic solid. \cite{guha2025complex} analyzed wave interactions in a graded piezomagnetic material, accounting for flexomagnetic, strain-gradient, and gravitational effects. \cite{biswas2024response,biswas2024plane} studied the reflection of plane waves at the boundary of flexomagnetic substrates .

In this work, we present a theory of elastic wave propagation in a linear elastic flexomagnetic solid with microstructure and gradient elasticity. In particular, for both longitudinal and transverse waves, we investigate the effect of flexomagnetism, microstructural length, and non-local interactions on the wave frequency, phase, and group velocities, and conditions for attenuation of waves, zero group velocity, negative group velocity, and the wave freezing phenomenon.

The remainder of this paper is organized as follows. In Section \ref{Sec:flexotheory}, we provide an outline of the theory of flexomagnetism in a linear elastic solid with microstructure. We use this theory to analyze the propagation of elastic waves in Section \ref{Sec:waves}, where we derive expressions for wave frequency, phase velocity, and group velocity for longitudinal and transverse waves in Sections \ref{Sec:Long} and \ref{Sec:Trans}, respectively, followed by a discussion on the effect of microstructure and magnetism on these quantities in Section \ref{Sec:discussion}. In Section \ref{Sec:Attenuation}, we discuss the range of wavenumbers over which elastic waves attenuate. In Section \ref{Sec:ZeroandNegativeGV}, we discuss the conditions for zero- and negative-group-velocity modes, and in Section \ref{Sec:WaveFreezing}, we analyze the wave-freezing phenomenon. Finally, in Section \ref{Sec:conclusion}, we provide concluding remarks. 

\section{Theory of flexomagnetism in an elastic solid with microstructure}\label{Sec:flexotheory}

Consider a linear elastic body occupying a region $\Omega$ in $\mathbb{R}^3$ and exhibiting flexomagnetism. The internal energy of the body is a function of the strain ${\bm{\varepsilon}}$, the gradient of strain $\nabla{\bm{\varepsilon}}$, the magnetization ${\bf{M}}$ and it's gradient $\nabla{{\bf{M}}}$  (\citep{Eliseev:PRB}), and is expressed as
{\begin{eqnarray}\label{Eq:IntEn}
  \mathcal{U}({{\bm{\varepsilon}},\nabla{\bm{\varepsilon}}, {\bf{M}}, \nabla{{\bf{M}}}}) &=& \frac{1}{2} C_{ijkl} u_{i,j}u_{k,l} + \frac{1}{2} h_{ijklmn} u_{i,jk} u_{l,mn} +\frac{1}{2} \mu_0 a_{kl}^m M_{k} M_{l} + \mu_0 d_{ijk}^m M_{i} u_{j,k} \nonumber \\ && +\frac{1}{2} \mu_0 g_{ijkl} M_{i,j} M_{k,l} + \mu_0 H_i^a M_i - \frac{1}{2} \mu_0 f_{lijk}^m ( u_{i,jk} M_l - u_{i,j} M_{l,k} ) \,\,\,,
\end{eqnarray}}
where the displacement field ${\bf{u}}$ and the strain ${\bm{\varepsilon}}$ are related by $\varepsilon_{ij} = \frac{1}{2} (u_{j,i} + u_{i,j}) $. In the Equation. \ref{Eq:IntEn}, the first term is due to linear elasticity and $C_{ijkl}$ is the fourth-rank elastic tensor. The second term is the high-order strain gradient elastic contribution which takes into account non-local behavior (\cite{mindlin1968first}), and $ h_{ijklmn}$ is a sixth-rank tensor. In the third term in (\ref{Eq:IntEn}), the second-rank tensor $a_{kl}^m$ is a material constant dependent on temperature. The fourth term in (\ref{Eq:IntEn}) is the piezomagnetic contribution, and $d_{ijk}^m$ is the third-rank piezomagnetic tensor. The tensor $g_{ijkl}$ is the inhomogeneous exchange coupling which incorporates the phenomenological continuum description of the quantum-mechanical exchange interaction. The next term is the interaction between the magnetization and an externally applied field $H_i^a$. This is also known as Zeeman energy (\cite{kronmuller2006}). Finally, the last term is the flexomagnetic contribution to the total internal energy, and $f_{ijkl}^m$ is the fourth-rank flexomagnetic coupling tensor. The flexomagnetic energy considered in this work is Lifschitz invariant, which considers both direct (i.e., magnetization from strain gradients) and converse (i.e., strain gradient from magnetization) flexomagnetic effects. In the present work, we neglect contributions from magnetocrystalline anisotropy (\cite{kronmuller2003micromagnetism}), magnetoelastic coupling (\cite{brown1966magnetoelastic}), and antisymmetric exchange interactions (\cite{dzyaloshinsky1958thermodynamic,moriya1960anisotropic}); however, the framework can be extended to incorporate these effects.

The body is assumed to have a microstructure, whose length-scale is $\lambda$, a real number (\cite{Mindlin1964}). The kinetic energy density of the system is due to the velocity as well as the inertia of the microstructure, which depends on velocity gradients (\cite{Mindlin1964,Papargyri2009})
{\begin{equation}\label{Eq:KE}
    T= \frac{\rho}{2} \dot{u}_i \dot{u}_i + \frac{1}{6} \rho \lambda^2 \dot{u}_{i,j} \dot{u}_{i,j} \,\,,
\end{equation}}
where $\rho$ is the mass density. We neglect the contributions of magnetization dynamics and its coupling with velocity to the total kinetic energy (\cite{Giannakopoulos2024,Yudin2013Review}).  

The constitutive relations for the total stress ${\bm{\sigma}}$ is obtained from the total internal energy using 
\begin{eqnarray} \label{Eq:stress1}
    {\bm{\sigma}}  &=& \frac{d  \mathcal{U} }{d \bm{\varepsilon}} =  \left[ \frac{\partial  \mathcal{U} }{\partial \bm{\varepsilon}}  - \nabla \cdot \frac{\partial  \mathcal{U} }{\partial \nabla  \bm{\varepsilon}} \right]   
\end{eqnarray} 
and the magnetic field strength ${\bf{H}}$ is
\begin{eqnarray}\label{Eq:H1}
    {\bf{H}} &=& \frac{1}{\mu_0}\frac{d \mathcal{U} }{d \bf{M}} =  \frac{1}{\mu_0}\left[ \frac{\partial \mathcal{U} }{\partial \bf{M}}  - \nabla \cdot \frac{\partial \mathcal{U} }{\partial \nabla  \bf{M}} \right]  \,\,.
\end{eqnarray} 
From these expressions, we obtain the stress as
\begin{equation}  \label{Eq:stress2}
    \sigma_{ij} = C_{ijkl} u_{k,l} + \mu_0d_{kij}^m M_k + \mu_0 f_{klij}^m M_{l,k} - h_{ijklmn} u_{l,mnk} 
\end{equation}
where the first two terms constitute the contribution of the Cauchy stress. The third and fourth terms represent the contributions from flexomagnetism and strain gradient elasticity, which include the higher-order stress contribution to the total stress tensor. For future reference, we denote the Cauchy stress and higher-order stress as 
\begin{eqnarray}
    \hat{\sigma}_{ij} &=&  C_{ijkl} u_{k,l} + \mu_0d_{kij}^m M_k \,\,, \nonumber \\
    \tau_{ijk} &=& -\mu_0 f_{klij}^m M_{l} + h_{ijklmn} u_{l,mn} \,\,,
\end{eqnarray}
such that $\sigma_{ij} =  \hat{\sigma}_{ij} - \tau_{ijk,k}$.

The magnetic field strength, using Equation \ref{Eq:H1}, is given by
\begin{equation} \label{Eq:H2}
    H_{k} = a_{kl}^m M_l + d_{kij}^m u_{i,j} + H_k^a - g_{ijkl} M_{i,jl} - f_{klij}^m u_{i,jl}  \,\,,
\end{equation} 
which comprises both the contributions from the local magnetic field strength and the higher-order magnetic field strength. These are
\begin{eqnarray}
    \hat{H}_{k} &=&   a_{kl}^m M_l + d_{kij}^m u_{i,j} + H_k^a \,\,,\nonumber\\
    S_{kl} &=& g_{ijkl} M_{i,j} + f_{klij}^m u_{i,j} \,\,,
\end{eqnarray}
such that $H_{k}=\hat{H}_{k}-S_{kl,l}$.


The total work done by external forces (\cite{georgiadis2004dispersive,gao2007variational,Papargyri2009,Hu:2018}) is
\begin{equation}
    \delta W = \int_{\Omega} \left( -\rho \ddot{u}_i + \frac{1}{3} \rho \lambda^2 \ddot{u}_{i,jj} + b_i \right) \delta u_i \,\, \mathrm{d\bm{x}}  +  \int_{\partial\Omega} \left( t_i - \frac{1}{3} \rho \lambda^2 \ddot{u}_{i,j} n_j + r_{i} \mathrm{D} \right)\delta u_i \mathrm{d\bm{s}}\label{varWork}
\end{equation}
where $b_i$ is the body force, $t_i$ is the Cauchy stress traction, and $r_i$ is the moment stress traction. In \ref{varWork}, the operator $\mathrm{D} = n_i\frac{\partial}{\partial x_i}$, and for future reference, the operator $D_i = (\delta_{ik}-n_in_k)\frac{\partial}{\partial x_i}$. Also, in (\ref{varWork}), the terms $\frac{1}{3} \rho \lambda^2 \ddot{u}_{i,jj}$ and $\frac{1}{3} \rho \lambda^2 \ddot{u}_{i,j} n_j$ are the micro-inertia force in the volume $\Omega$ and on the surface $\partial\Omega$, respectively (\cite{Hu:2018}). 

The free enthalpy of the system is
\begin{equation}\label{Eq:freeenthalpy}
    \mathcal{H}[u_i,H_i] = \int_{\mathbb{R}^3} \left( \mathcal{U} -\frac{1}{2} \mu_0 H_i H_i - \mu_0H_iM_i \right) \,\,\mathrm{d\bm{x}}  - W \,\,
\end{equation}
and the corresponding variational principle is
\begin{equation}\label{varPrinciple}
  \delta \mathcal{H} = \delta W \,\,,
\end{equation}
subject to the constraints
\begin{eqnarray}
    && \nabla \times {\bf{H}} = 0 \,\, \text{in} \,\, \Omega \,\,, \\
  \text{and} && \nabla \cdot {\bf{B}} = 0 \,\, \text{in} \,\, \Omega \,\,,
\end{eqnarray}
where ${\bf{B}}=\mu_0 ({\bf{H}} + {\bf{M}})$ is the magnetic field. As the magnetic field strength is curl-free, a magnetic scalar potential $\psi$ is defined as (\cite{coey2010magnetism, blundell2001magnetism})
\begin{equation}
    {\bf H} = -\nabla \psi \,\,.
\end{equation}

Taking variations of the free enthalpy Equation (\ref{Eq:freeenthalpy}), we obtain
\begin{eqnarray}\label{Eq:deltaH_1}
 \delta \mathcal{H} =  \int_{\mathbb{R}^3} \left( \frac{\partial \mathcal{U}}{\partial \varepsilon_{ij}} \delta \varepsilon_{ij} + \frac{\partial \mathcal{U}}{\partial M_k} \delta M_k + \frac{\partial \mathcal{U}}{\partial \varepsilon_{ij,k}} \delta \varepsilon_{ij,k} + \frac{\partial \mathcal{U}}{\partial M_{k,l}} \delta M_{k,l} - \mu_0 H_i \delta H_i -\mu_0 M_i \delta H_i - \mu_0 H_i \delta M_i\right) \mathrm{d\bm{x}}  
\end{eqnarray}
Upon substituting Equation (\ref{Eq:IntEn}) in Equation (\ref{Eq:deltaH_1}), we get
\begin{eqnarray}
 -\delta \mathcal{H} &=&  \int_{\mathbb{R}^3} \sigma_{ij,j} \delta u_i \,\mathrm{d\bm{x}}  +  \int_{\partial\Omega} \biggl( -\hat{\sigma}_{ij}n_j - (D_ln_l)\, n_jn_k \tau_{kji} + D_{j} \tau_{kji} n_k + \tau_{kji,k} n_j \biggr) \delta u_i \,\mathrm{d\bm{s}} \nonumber \\ && - \int_{\mathbb{R}^3} (H_i + \psi_{,i}) \delta M_i \,\, \mathrm{d\bm{x}}  - \int_{\partial \Omega} S_{ij} n_j \delta M_i \,\, \mathrm{d\bm{s}} + \int_{\mathbb{R}^3} \biggl( \mu_0 (- \psi_{,i} + M_i)_{,i} \biggr) \delta \psi \,\, \mathrm{d\bm{x}}  \nonumber \\
&& + \int_{\partial \Omega} \mu_0 (-\psi_{,i} + M_i) n_i \delta \psi \,\, \mathrm{d\bm{x}}  - \int_{\partial \Omega} (\tau_{kji} n_k n_j \,\mathrm{D} \delta u_i) \,\, \mathrm{d\bm{s}} \,\,. \label{Eq:var1}
\end{eqnarray}
From Equations \ref{varWork},  \ref{varPrinciple} and \ref{Eq:var1}, we obtain

\begin{eqnarray}\label{Eq:var2}
& & \int_{\mathbb{R}^3} \left( \sigma_{ij,j}  -\rho \ddot{u}_i + \frac{1}{3} \rho \lambda^2 \ddot{u}_{i,jj} + b_i \right) \delta u_i \,\mathrm{d\bm{x}}  \nonumber \\  && +  \int_{\partial\Omega} \left[ t_i -\hat{\sigma}_{ij}n_j -\frac{1}{3}\rho \lambda^2 \ddot{u}_{i,j} n_j - (D_sn_s) n_j n_k \tau_{kji} + D_{j} \tau_{kji} n_k + \tau_{kji,k} n_j \right] \delta u_i \,\mathrm{d\bm{s}} \nonumber \\ && - \int_{\mathbb{R}^3} (H_i + \psi_{,i}) \delta M_i \,\, \mathrm{d\bm{x}}  - \int_{\partial \Omega} S_{ij} n_j \delta M_i \,\mathrm{d\bm{s}} + \int_{\mathbb{R}^3} \left[ \mu_0(-\psi_{,i} + M_i)_{,i} \right] \delta \psi \,\mathrm{d\bm{x}}  \nonumber \\
&& + \int_{\partial \Omega} \mu_0 (-\psi_{,i} + M_i) n_i \delta \psi \,\mathrm{d\bm{x}}  - \int_{\partial \Omega} (r_i - \tau_{kji} n_k n_j) \mathrm{D} \delta u_i \,\mathrm{d\bm{s}} = 0\,\,.
\end{eqnarray}

From above Equation \ref{Eq:var2}, and neglecting body forces ($b_i=0$), we obtain the following field equations
\begin{eqnarray}\label{Eq:GovEqn}
     && \sigma_{ij,j} = \rho \ddot{u}_i - \frac{1}{3} \rho \lambda^2 \ddot{u}_{i,jj}  \,\,, \nonumber \\
     && H_{i} + \psi_{,i} = 0 \,\,, \nonumber \\
     && \mu_0 (-\psi_{,i} + M_i)_{,i} = 0 \,\, \text{in } \,\, \Omega \,\,, \nonumber \\
     \text{and}\,\, && \psi_{,ii} = 0 \,\, \text{in } \,\, \mathbb{R}^3 \setminus \Omega \,\,. 
\end{eqnarray}
The boundary conditions for the magnetic field and tractions along the smooth boundary $\partial\Omega$ are
\begin{eqnarray}\label{Eq:BC}
    \text{magnetic field strength:} && \mu_0 \left( -\psi_{,i} + M_i \right) n_i = 0 \,\,, \nonumber \\
     \text{higher-order magnetic field strength:} && S_{ij} n_j = 0 \,\,,  \nonumber \\
    \text{Cauchy stress traction:} && t_i = \hat{\sigma}_{ij} n_j + (D_sn_s) n_j n_k \tau_{kji} -D_j\tau_{kji,k} n_{k} + \frac{1}{3} \rho \lambda^2 \ddot{u}_{i,j} n_j \,\,, \nonumber \\
    \text{moment stress traction:} && r_i = \tau_{kji} n_k n_j \,\,. 
\end{eqnarray} 
Using the expression of stress (Equation \ref{Eq:stress2}) and magnetic field strength (Equation \ref{Eq:H2}), the set of field equations in (\ref{Eq:GovEqn}) are written in terms of the displacement field $u_i$ and magnetization $M_i$
\begin{eqnarray} \label{Eq:FieldEq}
     && C_{ijkl} u_{k,lj} + \mu_0 d_{kij}^m M_{k,j} + \mu_0 f_{klij}^m M_{l,kj} - h_{ijklmn} u_{l,mnkj} = \rho \ddot{u}_i - \frac{1}{3} \rho \lambda^2 \ddot{u}_{i,jj} \,\,, \nonumber \\
     && a_{ij}^m M_j + d_{ikl}^m u_{k,l} + H_i^a - g_{klij} M_{k,lj} - f_{klij}^m u_{k,lj} + \psi_{,i} = 0 \,\,, \nonumber \\
     && M_{i,i} -\psi_{,ii} = 0 \,\, \text{in } \,\, \Omega \,\,, \nonumber \\
     \text{and}\,\, && \psi_{,ii} = 0 \,\, \text{in } \,\, \mathbb{R}^3 \setminus \Omega \,\,. 
\end{eqnarray}
We assume the material to be isotropic. Then, $C_{ijkl}$, $h_{ijklmn}$, $f_{ijkl}^m$,  $g_{ijkl}$, $a_{ij}^m$ are given by (\cite{Hu:2018,Eliseev:PRB})
\begin{eqnarray}\label{Eq:tensor}
  &&   C_{ijkl} = c_{12} \delta_{ij} \delta_{kl} + c_{44}(\delta_{ik}\delta_{jl} + \delta_{il} \delta_{jk})   \,\,, \nonumber \\
  &&    h_{ijklmn} = l^2 C_{ijkl} \delta_{mn} =  c_{12} l^2 \delta_{ij} \delta_{kl}  \delta_{mn} + c_{44} l^2 (\delta_{ik}\delta_{jl} + \delta_{il} \delta_{jk})  \delta_{mn}  \,\,, \nonumber \\
  &&   f_{ijkl}^m =  f_{12}^m \delta_{ij} \delta_{kl} + f_{44}^m (\delta_{ik}\delta_{jl} + \delta_{il} \delta_{jk})\,\,, \nonumber \\
  && g_{ijkl} = A\,\delta_{ik} \delta_{jl} \,\,, \nonumber \\
 &&  a_{ij}^m = a_m \delta_{ij} \,\,,
\end{eqnarray} 
where, $c_{12}$, $c_{44}$ are the normal coupling elastic constant and the shear elastic constant, respectively, $f_{12}^m$ and $f_{44}^m$ are the longitudinal and shear flexomagnetic constant, respectively, $A$ is the magnetic exchange constant, $a_m$ is a material constant, and $l$ is a nonlocal elastic interaction length that quantifies the extent of non-local interactions in the linear theory of strain gradient elasticity (\cite{mindlin1951influence,mindlin1968first,Maranganti2006,Papargyri2009}) 

Using (\ref{Eq:tensor}), the field Equations \ref{Eq:FieldEq} are
\begin{eqnarray}
    && c_{44} u_{i,jj} + (c_{12}+c_{44}) u_{j,ji} + \mu_0 d_{kij}^m M_{k,j} + \mu_0 f_{44}^m M_{i,jj} + \mu_0 (f_{12}^m + f_{44}^m) M_{j,ji} - c_{44} l^2 u_{i,jjkk} - (c_{12} + c_{44}) l^2 u_{j,jikk} \nonumber \\ 
    && =  \rho \ddot{u}_i - \frac{1}{3} \rho \lambda^2 \ddot{u}_{i,jj} \,\,, \label{Eq:FieldEqLinElas1} \\
    && a_m M_i + d_{ikl}^m u_{k,l} - f_{44}^m u_{i,jj} - (f_{12}^m + f_{44}^m) u_{j,ji} + H_i^a + A M_{i,jj}^2 + \psi_{,i} = 0 \,\,, \label{Eq:FieldEqLinElas2} \\
    && M_{i,i} -\psi_{,ii} = 0  \label{Eq:FieldEqLinElas3} \,\,. 
\end{eqnarray}
The above equations can be solved for the field variables $u_i$, $M_i$, and $\psi$ to satisfy the boundary conditions provided in Equation \ref{Eq:BC}. We refer the reader to the works of \cite{hrytsyna2023love,jiao2024dispersion,biswas2024response,biswas2024plane,guha2025complex} for analysis and discussion of waves in piezomagnetic solids. In our present work, we neglect the effect of piezomagnetism (i.e., $d_{ikl}^m$ = 0) and focus our attention on the effects of flexomagnetism on waves in a linear elastic solid with microstructure and strain gradient elasticity. 

\section{Wave propagation in a flexomagnetic medium with microstructure}\label{Sec:waves}
\subsection{Longitudinal waves}\label{Sec:Long}

\begin{figure}[h!]
\centering
\includegraphics[keepaspectratio=true,width=0.65\textwidth]{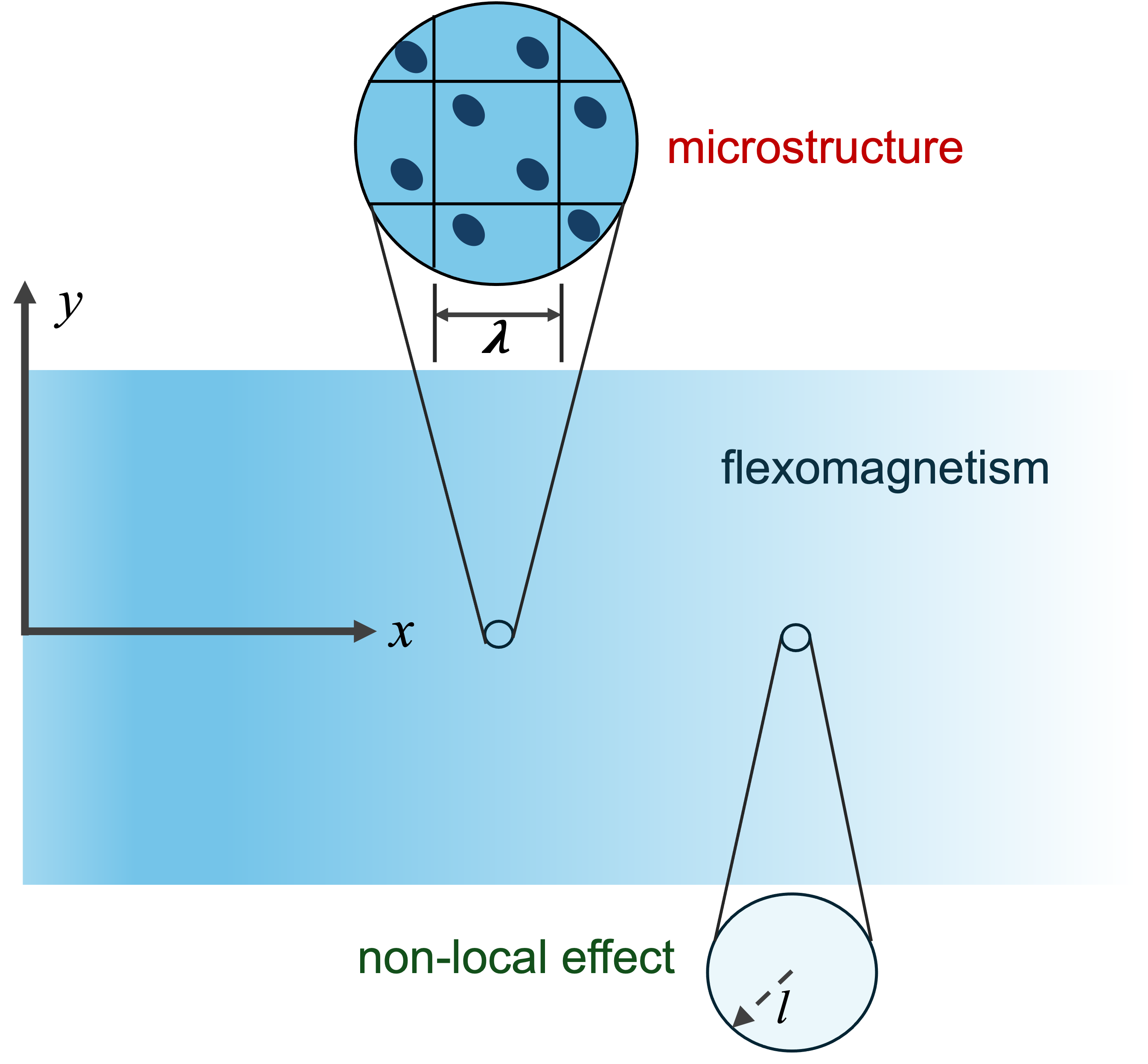}\label{Fig:schematic}
{\caption{A semi-infinite elastic solid with microstructure and non-local elastic interaction. The lengthscale of the microstructure is $\lambda$, and the non-local elastic interactions are restricted to a spherical region of radius $l$. The solid exhibits flexomagnetism. The wave is assumed to propagate in the $x$-direction.}}
\end{figure}

We now discuss the propagation of longitudinal waves in a semi-infinite isotropic elastic flexomagnetic solid, as shown in the Figure. \ref{Fig:schematic}. The direction of motion of longitudinal waves is in the direction of wave propagation. The field equations are described by (\ref{Eq:FieldEqLinElas1}- \ref{Eq:FieldEqLinElas3}). The applied field is absent, and the boundary conditions in Equation \ref{Eq:BC} are satisfied. The wave is assumed to propagate in the $x$-direction, and hence the field variables only depend on $x$. We denote $c=c_{12}+2c_{44}$ and $f_m = f_{12}^m+2f_{44}^m$, the field equations describing wave propagation are
\begin{eqnarray}\label{Eq:WaveEq1}
    && \mu_0 f_m \frac{\partial^2 M}{\partial x^2} + c \frac{\partial^2 u}{\partial x^2}  - c l^2 \frac{\partial^4 u}{\partial x^4} = \rho \frac{\partial ^2 u}{\partial t^2} -\frac{1}{3} \rho \lambda^2 \frac{\partial^4 u}{\partial x^2 \partial t^2} \,\,, \\
    && (a_m+1) M + A \left(\frac{\partial^2 M}{\partial x^2} \right)^2  - f_m \frac{\partial^2 u}{ \partial x^2}  = 0 \,\,. \label{Eq:WaveEq2}
\end{eqnarray}

We assume the following \emph{ansatz} for the displacement and magnetization to the above set of equations
\begin{eqnarray}\label{Eq:Sol}
     u(x,t) &=& u_0 \exp(i(kx-\omega t)) \,\,, \nonumber \\
   M(x,t) &=& m_0\exp(ikx)\mathcal{M}(t) \,\,, 
\end{eqnarray}
where $u$ and $M$ are in $x$-direction and  $\omega$ is the circular frequency and $k$ is the wave number and $i=\sqrt{-1}$. The function $\mathcal{M}(t)$ is the temporal dependence of magnetization, and can be calculated using the Landau-Lifschitz-Gilbert (LLG) equation (\cite{landau1935,Gilbert2004}).

Substituting (\ref{Eq:Sol}) in Equations \ref{Eq:WaveEq1} and \ref{Eq:WaveEq2}, we obtain the following set of simultaneous equations for $M(x,t)$ and $u(x,t)$
\begin{eqnarray}
 &&  \mu_0 k^2 f_m M(x,t) + c k^2 u(x,t) + c l^2 k^4 u(x,t) = \rho \omega^2 u(x,t) + \frac{1}{3} \rho \lambda^2 k^2 \omega^2 u(x,t) \,\,, \label{Eq:F1:L} \\ 
 &&   A k^4 M(x,t)^2 + (a_m+1) M(x,t) + f_m k^2 u(x,t) = 0 \,\,. \label{Eq:F2:L}
\end{eqnarray}

Restricting our attention to  materials with small exchange stiffness, we solve Equations \ref{Eq:F1:L}, and \ref{Eq:F2:L}, and obtain 
\begin{equation}
   M(x,t) = - \frac{k^2 f_m}{a_m + 1} u(x,t) \,\,,
\end{equation}
and
\begin{eqnarray}\label{Eq:omegaEqn2}
\frac{k^4 \mu_0 f_m^2}{a_m+1}-c \left(k^4 l^2+k^2\right)+\frac{1}{3} \rho  \omega ^2 \left(k^2 \lambda ^2+3\right) = 0 \,\,.
\end{eqnarray}


Solving Equation \ref{Eq:omegaEqn2}, for the circular frequency of the longitudinal wave
\begin{equation}\label{Eq:omegal}
    \omega^l = k \left[ \frac{c(a_m+1)(k^2l^2+1) - \mu_0 k^2 f_m^2}{\rho (a_m+1) (\frac{1}{3}k^2 \lambda^2 + 1)} \right]^{\frac{1}{2}}
\end{equation}
The phase velocity of the longitudinal wave is $V_p^l = \omega^l/k$. 

From Equation \ref{Eq:omegal}, we see that when microstructure, non-local elastic interactions, and magnetism, are absent i.e. ($\lambda=0, l=0, a_m=0, f_m=0$), the \emph{classical phase velocity} of a longitudinal wave propagating in a linear elastic medium and is given by $V_p^{c,l} = \sqrt{c/\rho}$ (\cite{Achenbach}). Also, from Equation \ref{Eq:omegal}, notice that in the absence of flexomagnetism (i.e. $f_m=0$), the circular frequency is given by
\begin{equation}\label{Eq:omegal2}
     \omega^l = k \left[ \frac{c(k^2l^2+1)}{\rho (\frac{1}{3}k^2 \lambda^2 + 1)} \right]^{\frac{1}{2}}\,\,,
\end{equation}
which is the expression derived by \cite{Papargyri2009}.

The group velocity of the longitudinal wave is given by the partial derivative of the circular frequency $\omega^l$ with respect to the wavevector $k$.

\begin{equation}\label{Eq:Vgl}
    V_g^l = \frac{\partial \omega^l}{\partial k} = \frac{\bigg(c(a_m+1)l^2 - \mu_0 f_m^2\biggr)\biggl( \frac{1}{3} \lambda^2 k^4 + 2 k^2\biggr)+c(a_m+1)}{\rho  \left(a_m+1\right) \left( \frac{1}{3} k^2 \lambda ^2+1\right)^2 \sqrt{\frac{c \left(a_m+1\right) \left(k^2 l^2+1\right)-k^2 \mu _0 f_m^2}{\rho  \left(a_m+1\right) \left( \frac{1}{3} k^2 \lambda ^2+1\right)}}} \,\,.
\end{equation}
In the absence of microstructure, non-local elastic interactions, and magnetism, the group velocity is equal to the classical group velocity of the longitudinal wave in a linear elastic medium $V_g^{c,l} = \sqrt{\frac{c}{\rho}}$ (\cite{Achenbach}), equal to the classical phase velocity of longitudinal waves. When flexomagnetic effects are absent, we substitute $f_m=0$ in Equation \ref{Eq:Vgl} and obtain the group velocity as 

\begin{equation}
    V_g^l  = \frac{\sqrt{\frac{c}{\rho}}\biggl\{ (\frac{1}{3} \lambda^2 k^4 + 2 k^2)  l^2+1\biggr\}}{\left( \frac{1}{3} k^2 \lambda ^2+1\right)^{3/2} \sqrt{{ \left(k^2 l^2+1\right)}}} \,\,.
\end{equation}

In the limit of small wavenumber i.e. $k\to 0$ and large wavenumber $k \to \infty$, the phase velocities and the group velocities are 
\begin{eqnarray}
    \lim_{k \to 0 } V_p^l &=& \sqrt{\frac{c}{\rho}} \,\,, \\
    \lim_{k \to \infty } V_p^l &=& \sqrt{\frac{c(a_m+1)l - \mu_0f_m^2}{\rho(a_m+1)\left( \frac{\lambda^2}{3}\right)}} \,\,.
\end{eqnarray}
\begin{eqnarray}
    \lim_{k \to 0 } V_g^l &=& \sqrt{\frac{c}{\rho}}  \,\,, \\
    \lim_{k \to \infty } V_g^l &=&  \sqrt{\frac{c(a_m+1)l - \mu_0f_m^2}{\rho(a_m+1)\left( \frac{\lambda^2}{3}\right)}} \,\,.
\end{eqnarray}
Notice from these expressions that in the limit of $k\to 0$, the phase velocity is independent of the length-scale parameters $\lambda$ and $l$, and coincides with the classical phase velocity. On the other hand, when $k \to \infty$, the phase velocity depends on both the microstructural length-scale parameter $\lambda$ and the nonlocal elastic interaction length $l$, accounting for non-local elastic interactions. Also, in the limit of $k\to 0$, the group velocity of longitudinal waves is independent of the length-scale parameters $\lambda$ and $l$, and is equal to the classical phase velocity of longitudinal waves. When $k \to \infty$, the group velocity is equal to the phase velocity. 

\subsection{Transverse waves}\label{Sec:Trans}
We discuss the propagation of transverse waves in a semi-infinite isotropic elastic flexomagnetic solid. The geometry is shown in Figure \ref{Fig:schematic}. The direction of motion of a transverse wave is orthogonal to the direction of wave propagation. We consider waves propagating along $x$ direction with unit vector $\hat{\bf{e}}_1$ and the wave amplitude is in the direction $\hat{\bf{n}}$ such that ${\hat{\bf{n}}} \cdot {\hat{\bf{e}}}_1 = 0$. Thus, we can write 
\begin{equation}
    u_{j,ji} = 0 \,\,, u_{j,jikk} = 0 \,\,, \text{and} \,\, M_{j,ji} = 0 \,\,.
\end{equation}
The field Equations \ref{Eq:FieldEqLinElas1} - \ref{Eq:FieldEqLinElas3} for transverse waves propagating in a linear elastic flexomagnetic solid without piezomagnetic effects and zero applied field are
\begin{eqnarray}\label{Eq:FieldEq2}
    && c_{44} u_{i,jj}  + \mu_0 f_{44}^m M_{i,jj} - c_{44} l^2 u_{i,jjkk} =  \rho \ddot{u}_i - \frac{1}{3} \rho \lambda^2 \ddot{u}_{i,jj} \,\,, \\
    && a_m M_i  - f_{44}^m u_{i,jj} + A M_{i,jj}^2 + \psi_{,i} = 0 \,\,, \\
    && M_{i,i} -\psi_{,ii} = 0  \,\,.
\end{eqnarray}

We assume the following ansatz for the displacement field and magnetization 
\begin{eqnarray}\label{Eq:Sol:T}
     u(x,t) &=& u_0 \exp(i(kx-\omega t)) \,\,, \nonumber \\
   M(x,t) &=& m_0\exp(ikx)\mathcal{M}(t) \,\,, 
\end{eqnarray}
where $u$ and $M$ are along a direction normal to the $x$-direction, $\omega$ is the circular frequency and $k$ is the wave number and $i=\sqrt{-1}$. The function $\mathcal{M}(t)$ is the temporal dependence of magnetization. We restrict ourselves to materials with low exchange stiffness. Substituting this ansatz in Equation \ref{Eq:FieldEq2}, we obtain
\begin{eqnarray} \label{Eq:omegaEqn2trans}
 &&  \mu_0 k^2 f_{44}^m M(x,t) + c_{44} k^2 u(x,t) + c_{44} l^2 k^4 u(x,t) = \rho \omega^2 u(x,t) + \frac{1}{3} \rho \lambda^2 k^2 \omega^2 u(x,t) \,\,, \label{Eq:F1:T} \\ 
 &&  (a_m+1) M(x,t) + f_{44}^m k^2 u(x,t) = 0 \,\,, \label{Eq:F2:T}
\end{eqnarray}

Solving Equation \ref{Eq:omegaEqn2trans}, the circular frequency of a transverse wave is
\begin{equation}\label{Eq:omegat}
    \omega^t = k \left[ \frac{c_{44}(a_m+1)(k^2l^2+1) - \mu_0(kf_{44}^m)^2}{\rho (a_m+1) (\frac{1}{3}k^2 \lambda^2 + 1)} \right]^{\frac{1}{2}}
\end{equation}
The phase velocity of a transverse wave is $V_p^t = \omega^t/k$. From Equation \ref{Eq:omegat}, we see that in the absence of microstructure, non-local elastic interactions, and magnetism i.e. ($\lambda=0, l=0, a_m=0, f_m=0$), we obtain the \emph{classical phase velocity} of a transverse wave propagating in a linear elastic medium as $V_p^{c,t} = \sqrt{c_{44}/\rho}$ (\cite{Achenbach}). 

Also from Equation \ref{Eq:omegat}, we see that flexomagnetism is absent (i.e. $f_m=0$), the circular frequency is given by
\begin{equation}\label{Eq:omegat2}
     \omega^t = k \left[ \frac{c_{44}(k^2l^2+1)}{\rho (\frac{1}{3}k^2 \lambda^2 + 1)} \right]^{\frac{1}{2}} \,\,,
\end{equation}
which is also the expression derived by \cite{Papargyri2009}.
The group velocity of a transverse wave is given by the partial derivative of the circular frequency $\omega^t$ with respect to the wavenumber

\begin{equation}\label{Eq:Vgt}
    V_g^t = \frac{\partial \omega^t}{\partial k} = \frac{\biggl(c_{44}(a_m+1)l^2 - \mu_0 (f^m_{44})^2\biggr)( \frac{1}{3} \lambda^2 k^4 + 2 k^2)+c_{44}(a_m+1)}{\rho  \left(a_m+1\right) \left( \frac{1}{3} k^2 \lambda ^2+1\right)^2 \sqrt{\frac{c_{44} \left(a_m+1\right) \left(k^2 l^2+1\right)-k^2 \mu _0 (f^m_{44})^2}{\rho  \left(a_m+1\right) \left( \frac{1}{3} k^2 \lambda ^2+1\right)}}} \,\,.
\end{equation}
When microstructure, non-local elastic interactions, and magnetism are absent, then the group velocity is equal to the classical group velocity of the transverse wave in a linear elastic medium $V_g^{c,t} = \sqrt{\frac{c_{44}}{\rho}}$ (\cite{Achenbach}), equal to the classical phase velocity of transverse waves. When flexomagnetic effects are absent, we substitute $f_m=0$ in Equation \ref{Eq:Vgt} and obtain the group velocity as 

\begin{equation}
    V_g^t  = \frac{\sqrt{\frac{c_{44}}{\rho}}\biggl\{ (\frac{1}{3} \lambda^2 k^4 + 2 k^2)  l^2+1\biggr\}}{\left( \frac{1}{3} k^2 \lambda ^2+1\right)^{3/2} \sqrt{{ \left(k^2 l^2+1\right)}}} \,\,.
\end{equation}

The phase and group velocities in the limit of small wavenumber i.e. $k\to 0$ and large wavenumber $k \to \infty$ are given by
\begin{eqnarray}
    \lim_{k \to 0 } V_p^t &=& \sqrt{\frac{c_{44}}{\rho}} \,\,, \\
    \lim_{k \to \infty } V_p^t &=& \sqrt{\frac{c_{44}(a_m+1)l - \mu_0(f_{44}^m)^2}{\rho(a_m+1)\left( \frac{\lambda^2}{3}\right)}} \,\,.
\end{eqnarray}
\begin{eqnarray}
    \lim_{k \to 0 } V_g^t &=& \sqrt{\frac{c_{44}}{\rho}}  \,\,, \\
    \lim_{k \to \infty } V_g^t &=&  \sqrt{\frac{c_{44}(a_m+1)l - \mu_0(f_{44}^m)^2}{\rho(a_m+1)\left( \frac{\lambda^2}{3}\right)}} \,\,.
\end{eqnarray}
It is evident from these expressions that when $k\to 0$, the phase and group velocities of transverse waves are independent of length parameters $\lambda$ and $l$ and are equal to the classical phase velocity. When $k \to \infty$, the group and phase velocities are equal, and depend on both the microstructural length parameter $\lambda$ and the nonlocal elastic interaction length $l$.

\subsection{Discussion on frequencies and velocities of longitudinal and transverse waves}\label{Sec:discussion}
In this section, we discuss the effect of flexomagnetism, the nonlocal elastic interaction length $l$, and the microstructural length $\lambda$ on the frequency and the phase and group velocities of longitudinal and transverse waves propagating in a semi-infinite flexomagnetic solid. We use the expressions derived in Sections \ref{Sec:Long} and \ref{Sec:Trans}. Table \ref{parametertable} shows the parameters used for our discussion.

\begin{table}[!h]
\centering
\caption{Material constants used in our study}
\label{parametertable}
\begin{tabular}{ll}
\hline
name & value \\
\hline
$c_{12}$ & 160 GPa (\cite{wang2012first}) \\
$c_{44}$ & 160 GPa (\cite{wang2012first}) \\
$c$ & 480 GPa \\
$\rho$ & 5220 Kg/m$^3$ (\cite{makushko2022}) \\
$\mu_0$ & $1.256637 \times 10^{-6}$ N/A$^2$ (\cite{sidhardh2018}) \\
$a_m$ & $125$ (\cite{sidhardh2018}) \\
$f_m$ &  $10^{-4}$ A \cite{Eliseev:PRB} \\
$f_{12}^m$ & $0.34\times10^{-4}$ A \\
$f_{44}^m$ & $0.33\times10^{-4}$ A \\
$\lambda$ & 0 - 40 Angstrom\\
$l$ & 0 - 100 Angstrom \\
\hline
\end{tabular}
\vspace*{-4pt}
\end{table}

\begin{figure}[h]
\centering
\subfigure[longitudinal]{\includegraphics[keepaspectratio=true,width=0.45\textwidth]{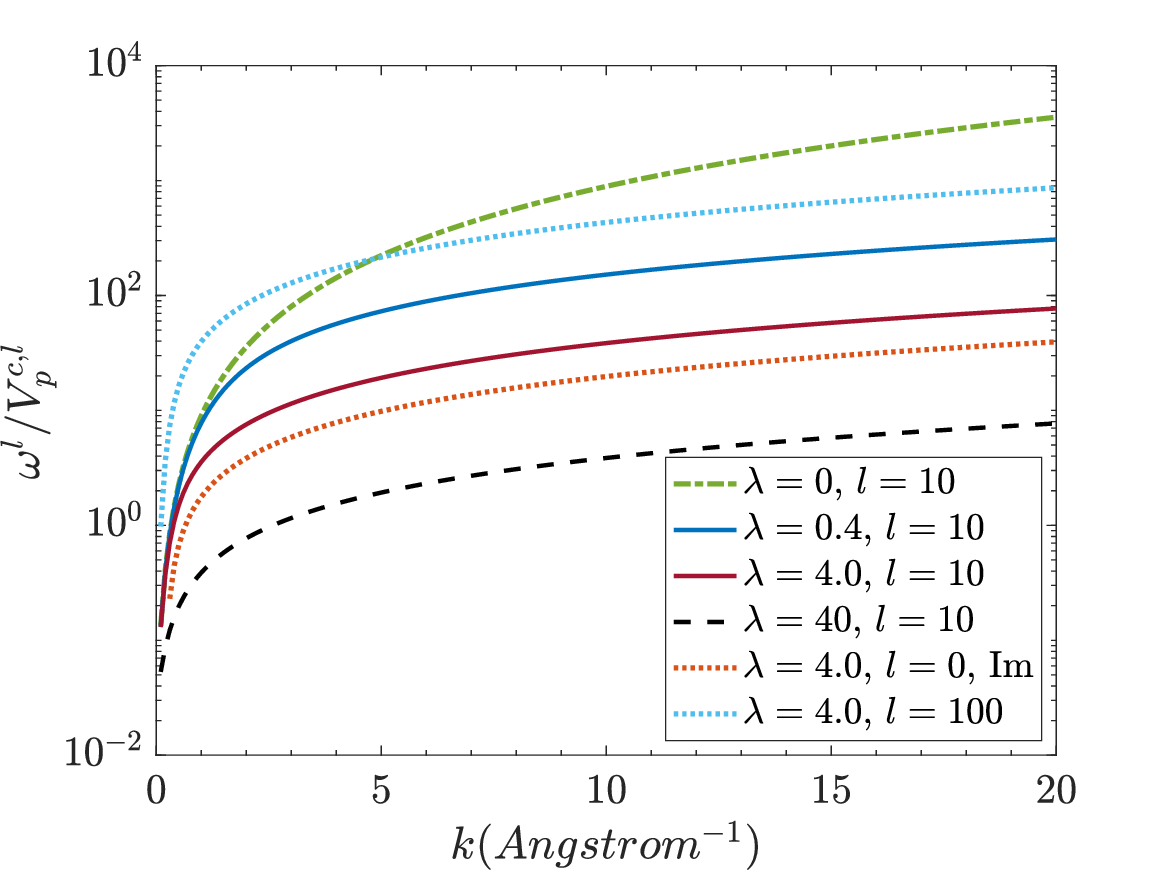}\label{Fig:OmegaL}}
\subfigure[transverse]{\includegraphics[keepaspectratio=true,width=0.45\textwidth]{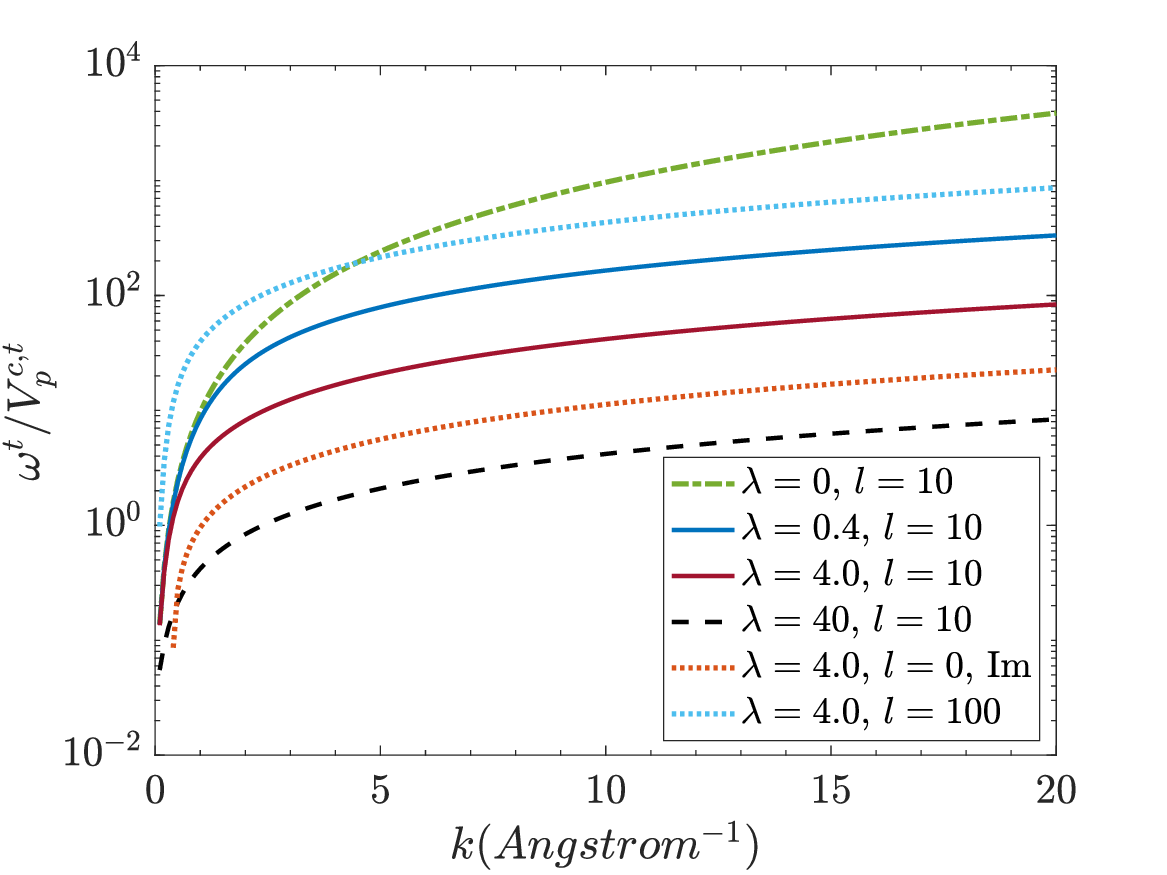}\label{Fig:OmegaT}}
{\caption{The effect of the length parameters $\lambda$ and $l$ on the wave frequency is shown for different values of wave number $k$. (a) shows the longitudinal case, and (b) shows the transverse case. The wave frequencies are normalized by the classical phase velocities of the longitudinal and transverse cases, respectively.}\label{Fig:Omega}}
\end{figure}

\paragraph{Wave frequency:} Figure \ref{Fig:Omega} shows the effect of micro-structural length $\lambda$ and the nonlocal elastic interaction length $l$ on the wave frequency $\omega$ vs wave number $k$ (dispersion relation) for longitudinal and transverse cases. From this figure, we see that keeping $l$ constant, if $\lambda$ is increased (or decreased), the wave frequency decreases (or increases). This corroborates with $\omega \sim \frac{1}{\lambda}$ as seen from Equations \ref{Eq:omegal2} and \ref{Eq:omegat2}. However, keeping $\lambda$ constant, when $l$ is increased (or decreased) then the wave frequency increases (or decreases), as $\omega \sim l$ in Equations \ref{Eq:omegal2} and \ref{Eq:omegat2}. 

Notice from Equations  \ref{Eq:omegal2} and \ref{Eq:omegat2}, that if $l=0$ (i.e. elastic interactions are purely local), then for longitudinal waves, $\omega^l$ is real for $|k| \leq \left[ \frac{c(a_m+1)}{\mu_0 f_m^2}\right]^{1/2}$ and for transverse waves, $\omega^t$ is real for $|k| \leq \left[ \frac{c_{44}(a_m+1)}{\mu_0 (f_{44}^m)^2}\right]^{1/2}$. When $\lambda=0$ (i.e. microstructure is absent), then for longitudinal waves, $\omega^l$ is real for $|k| \leq \left[ \frac{c(a_m+1)}{\mu_0 f_m^2 - c(a_m+1)l^2}\right]^{1/2}$ and for transverse waves , $\omega^t$ is real for $|k| \leq \left[ \frac{c_{44}(a_m+1)}{\mu_0 (f_{44}^m)^2- c_{44}(a_m+1)l^2}\right]^{1/2}$.

\begin{figure}[h]\centering
\subfigure[longitudinal]{\includegraphics[keepaspectratio=true,width=0.4\textwidth]{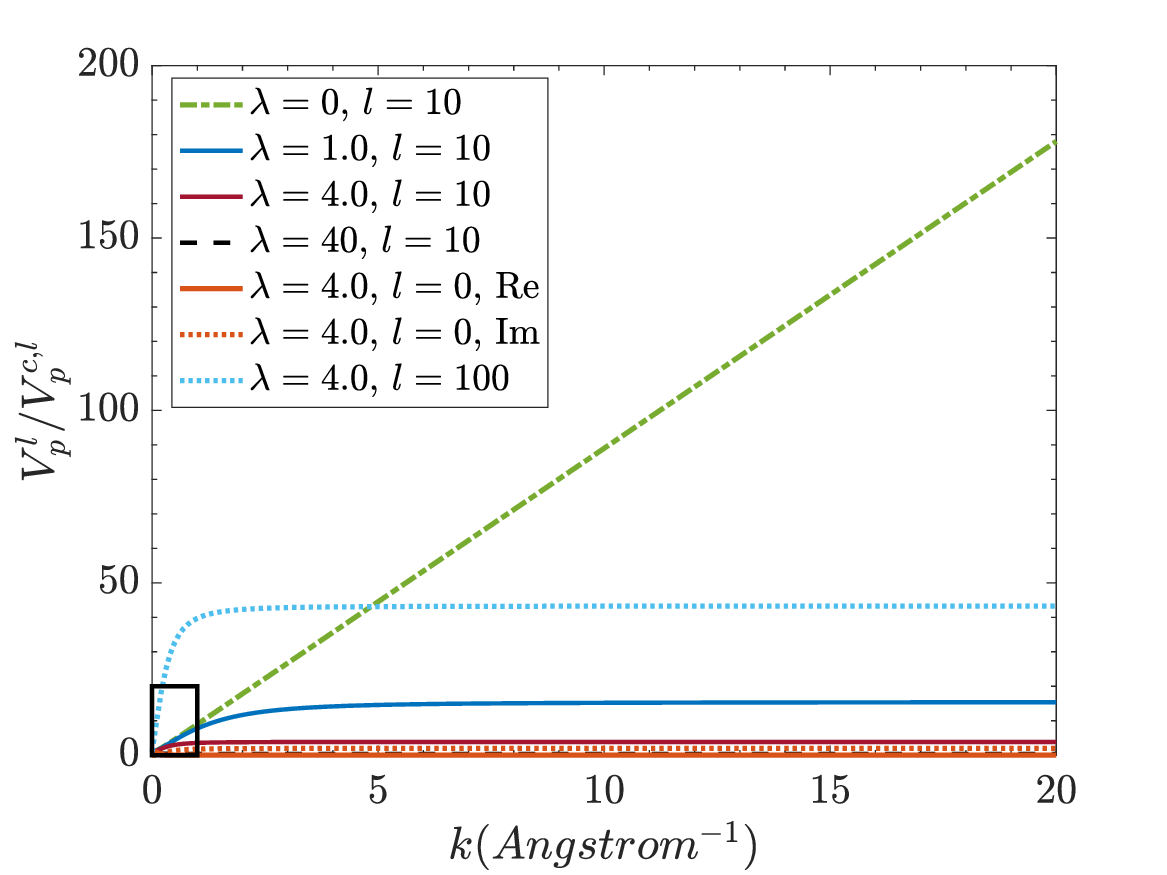}}
\subfigure[longitudinal]{\includegraphics[keepaspectratio=true,width=0.4\textwidth]{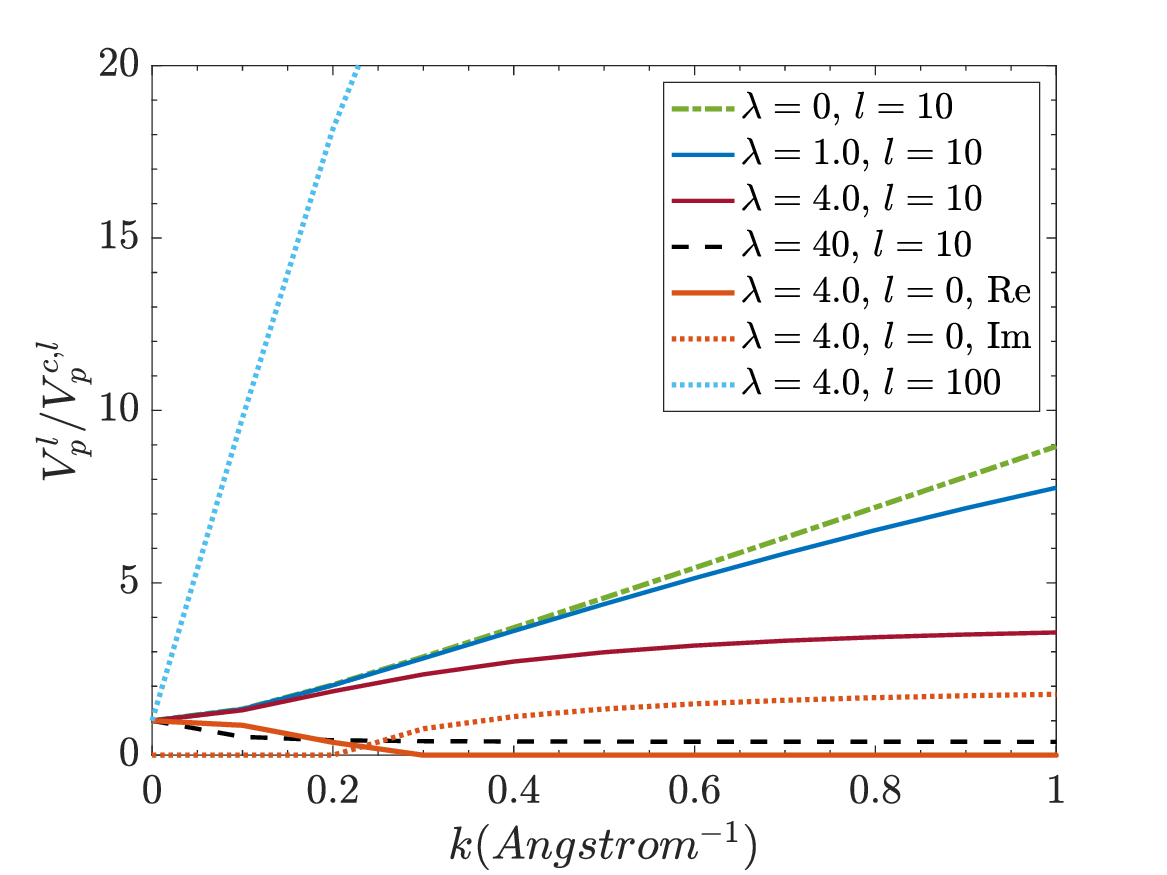}} \\
\subfigure[transverse]{\includegraphics[keepaspectratio=true,width=0.4\textwidth]{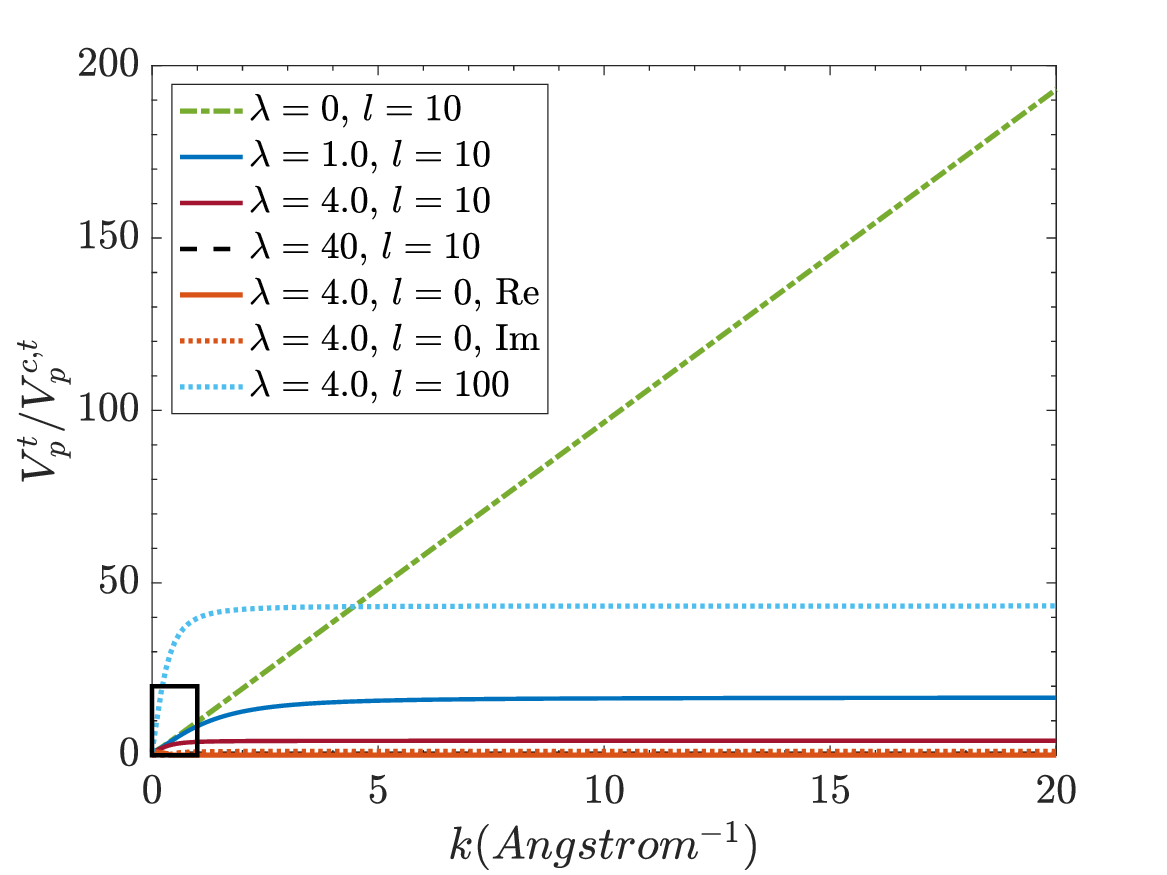}}
\subfigure[transverse]{\includegraphics[keepaspectratio=true,width=0.4\textwidth]{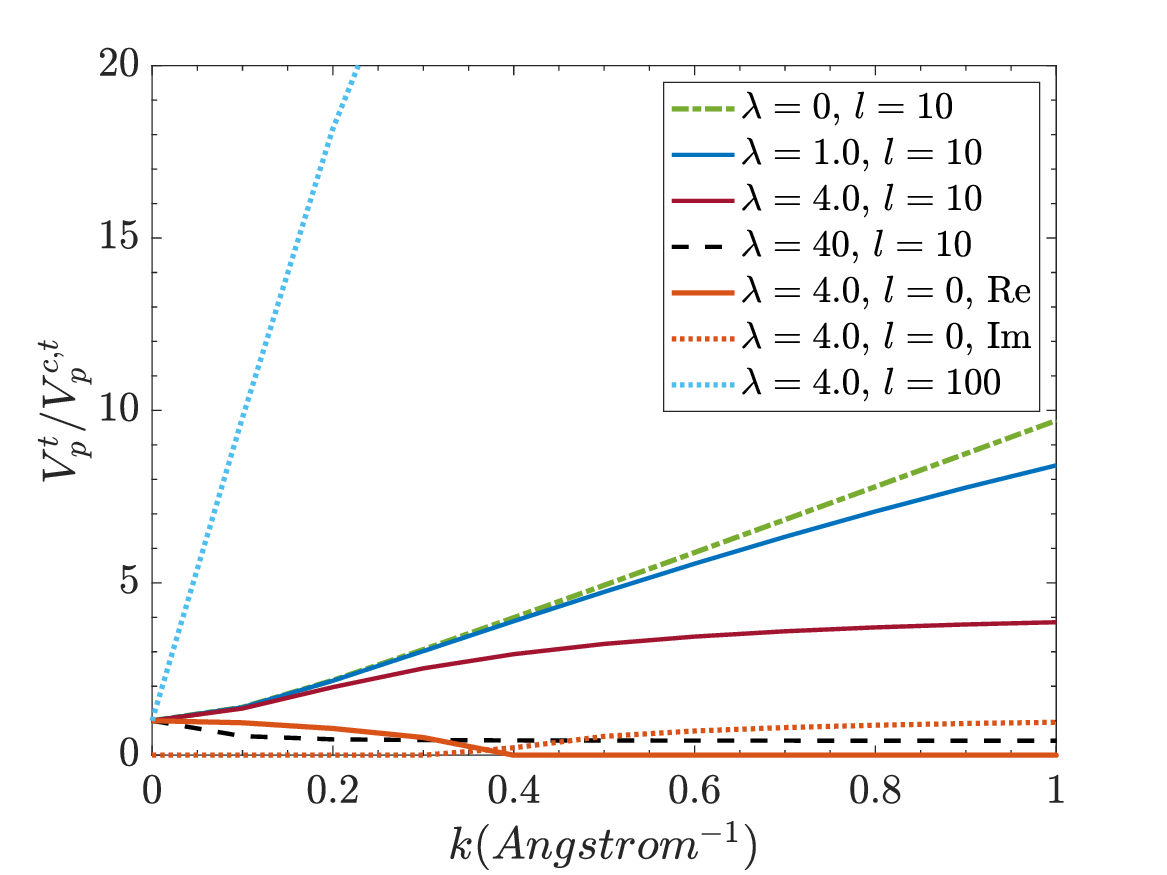}} 
{\caption{The effect of length parameters $\lambda$ and $l$ on the phase velocities of (a)-(b) longitudinal and (c)-(d) transverse waves is shown for different values of wave number $k$. The phase velocities are normalized by the classical phase velocities. (b) and (d) shows an inset of (a) and (c), respectively, from wave numbers 0 to 1 Angstrom$^{-1}$, denoted by the black box in (a) and (c).}\label{Fig:VpVpc}}
\end{figure}

\paragraph{Phase velocity:}
Figure \ref{Fig:VpVpc} shows the effect of non-local interaction and microstructure on the normalized phase velocity versus wave number curves for both longitudinal and transverse cases. From these plots, we see that when $\lambda=0$ and $l=10$ Angstrom, the phase velocity increases unboundedly and is not physical. keeping $l=10$ Angstrom  and increasing the microstructural length parameter $\lambda$ to $1$, $4.0$, and $40.0$ Angstroms, the phase velocities decrease with increasing $\lambda$ and is bounded, hence physically meaningful. When the nonlocal elastic interaction length $l$ is set to zero, and $\lambda$ = 4.0 Angstrom, then the phase velocity is real for $k\leq \left[ \frac{c(a_m+1)}{\mu_0 f_m^2}\right]^{1/2}$ = 0.2194 Angstrom$^{-1}$ for longitudinal waves and $k \leq \left[ \frac{c_{44}(a_m+1)}{\mu_0 (f_{44}^m)^2}\right]^{1/2}$ = 0.3838 Angstrom$^{-1}$ for transverse waves. In Figure \ref{Fig:VpVpc}, we plot both the real and imaginary components of this phase velocity. The imaginary part of the phase velocity is bounded while the real part is zero with increasing wavenumbers. When $l$ is increased to 100 Angstrom, the phase velocity is bounded. Overall, we see that the microstructural and nonlocal elastic interaction length-scale parameters have contrasting effects on the phase velocity. While increasing $\lambda$ decreases the phase velocity, increasing $l$ increases the phase velocity.

\begin{figure}[h]\centering
\subfigure[longitudinal]{\includegraphics[keepaspectratio=true,width=0.45\textwidth]{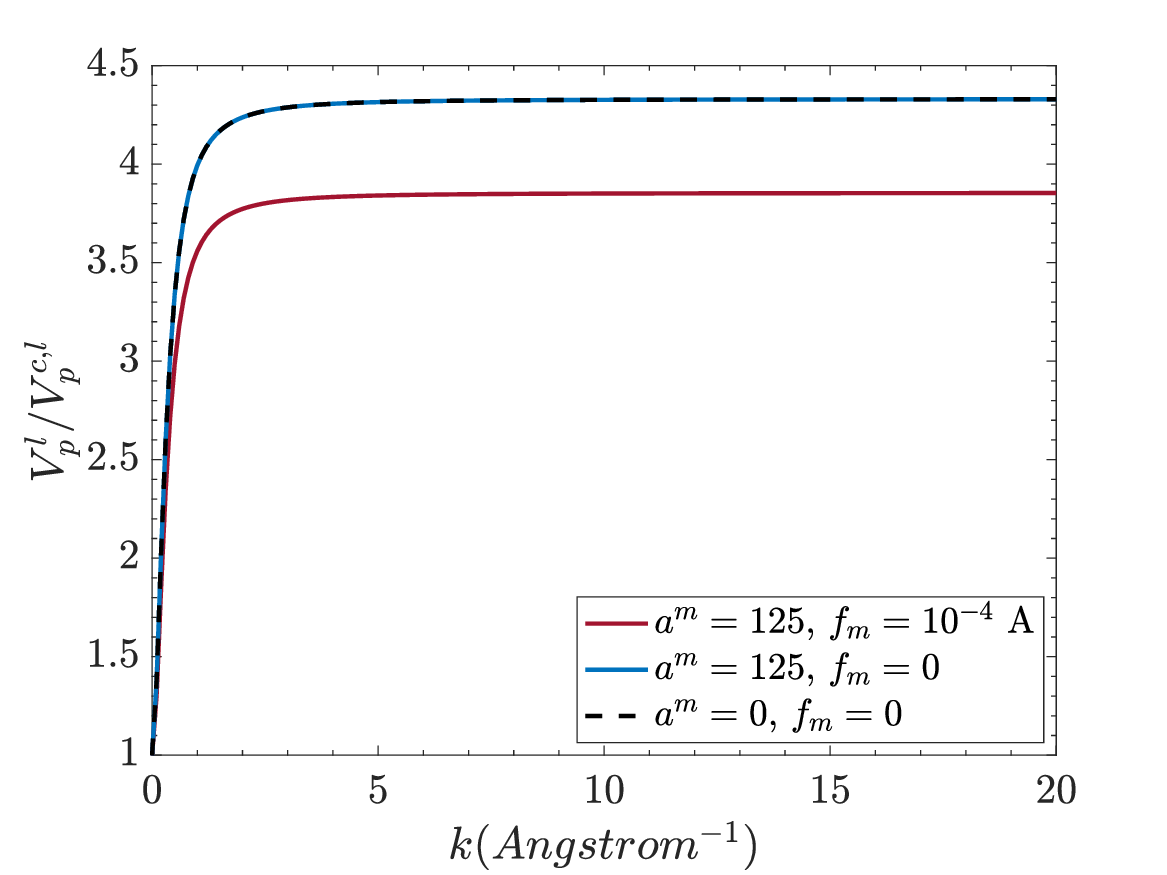}}
\subfigure[transverse]{\includegraphics[keepaspectratio=true,width=0.45\textwidth]{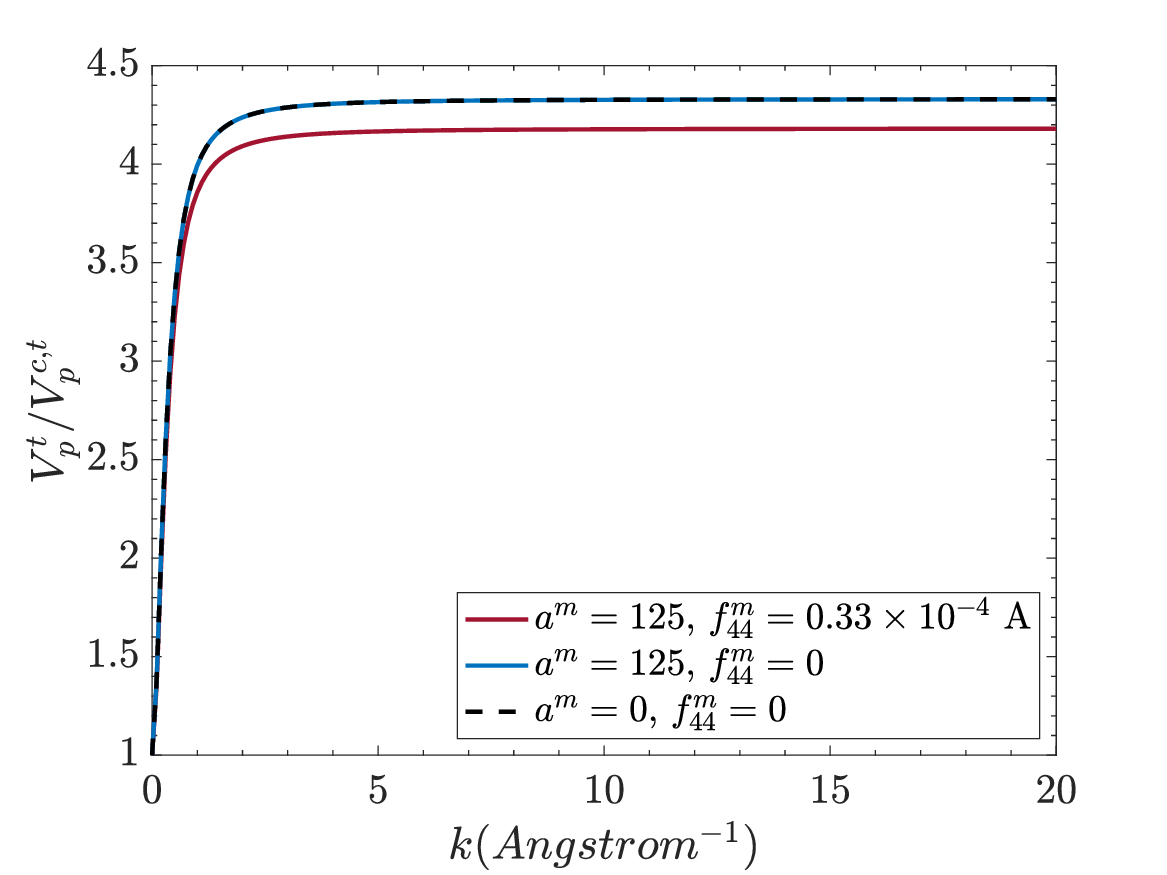}}
{\caption{The effect of magnetism on the phase velocities of (a) longitudinal and (b) transverse waves is shown for different values of wave number $k$. The phase velocities are normalized by the classical phase velocities.}\label{Fig:VpVpcMag}}
\end{figure}

In Figure \ref{Fig:VpVpcMag}, we show effect of magnetism on the phase velocities of longitudinal and transverse waves. We choose $\lambda$ = 4.0 Angstrom and $l$ = 10 Angstrom for this comparison. Figure \ref{Fig:VpVpcMag} shows that flexomagnetic effects decrease the phase velocity. Also, the phase velocity in a non-magnetic material is the same as that in a magnetic material without flexomagnetism. This can be seen by setting $f^m$ = 0 and $f_{12}^m$ = 0 in Equations \ref{Eq:omegal} and \ref{Eq:omegat}, respectively. Furthermore, as the phase velocity of longitudinal waves depends on $f_{m}$ whereas the phase velocity of transverse waves depends on $f_{44}^m$ and $f_{m} > f_{44}^m$ the influence of flexomagnetism on the phase velocity of longitudinal waves is more than that for transverse waves.

\begin{figure}[h]\centering
\subfigure[longitudinal]{\includegraphics[keepaspectratio=true,width=0.45\textwidth]{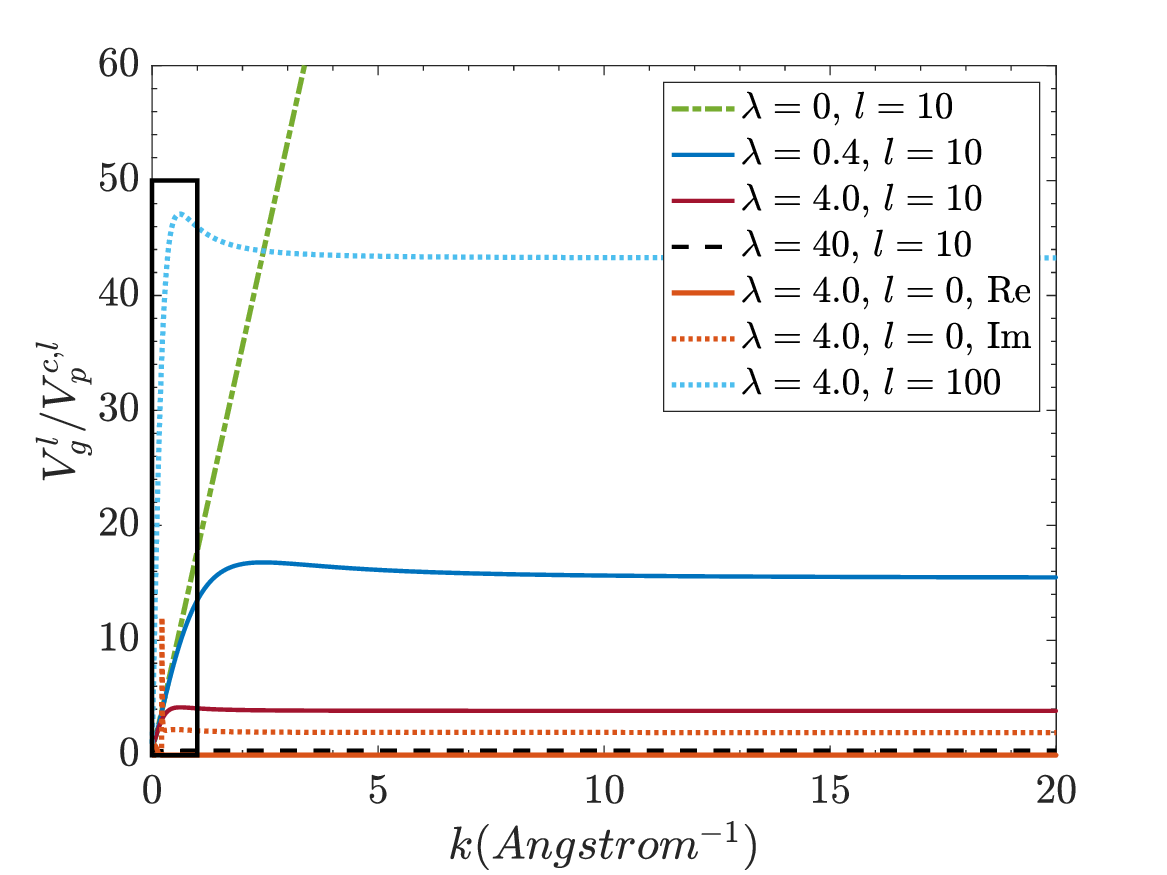}}
\subfigure[longitudinal]{\includegraphics[keepaspectratio=true,width=0.45\textwidth]{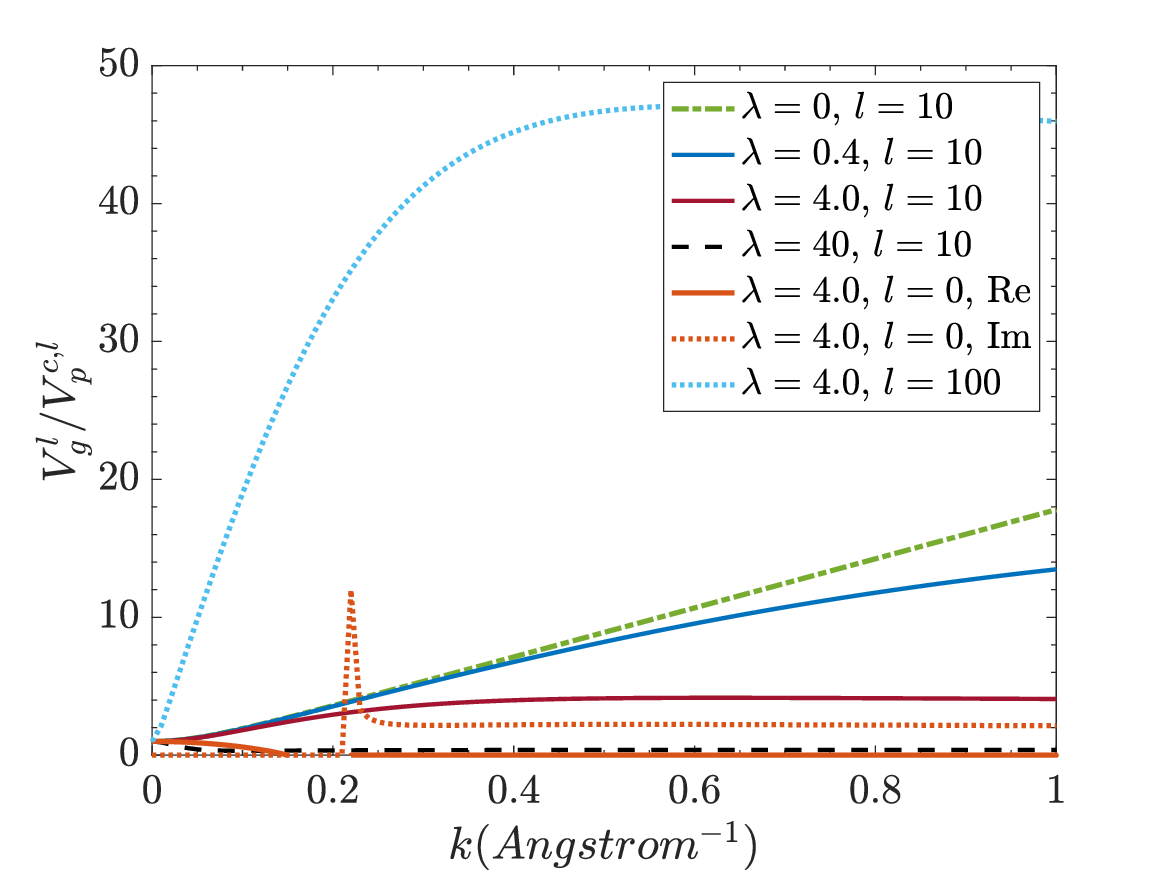}}\\
\subfigure[transverse]{\includegraphics[keepaspectratio=true,width=0.45\textwidth]{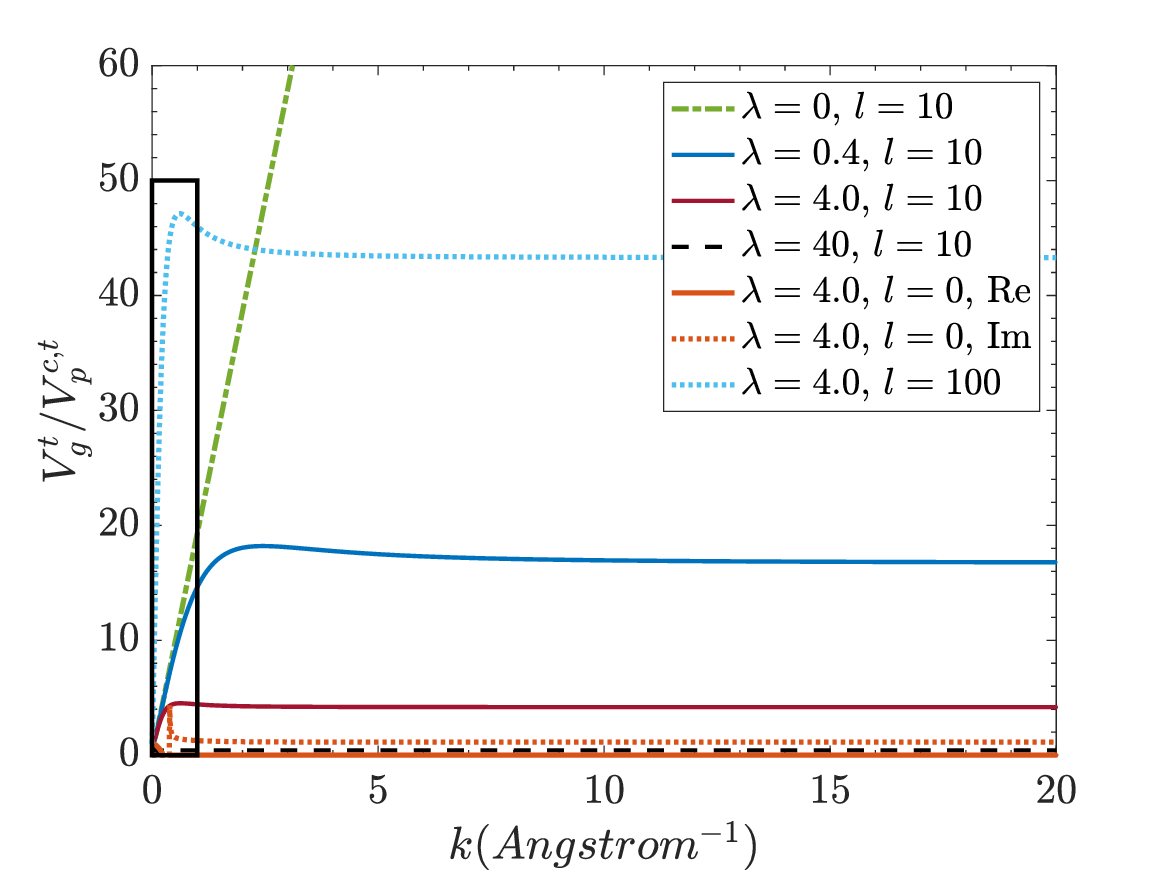}}
\subfigure[transverse]{\includegraphics[keepaspectratio=true,width=0.45\textwidth]{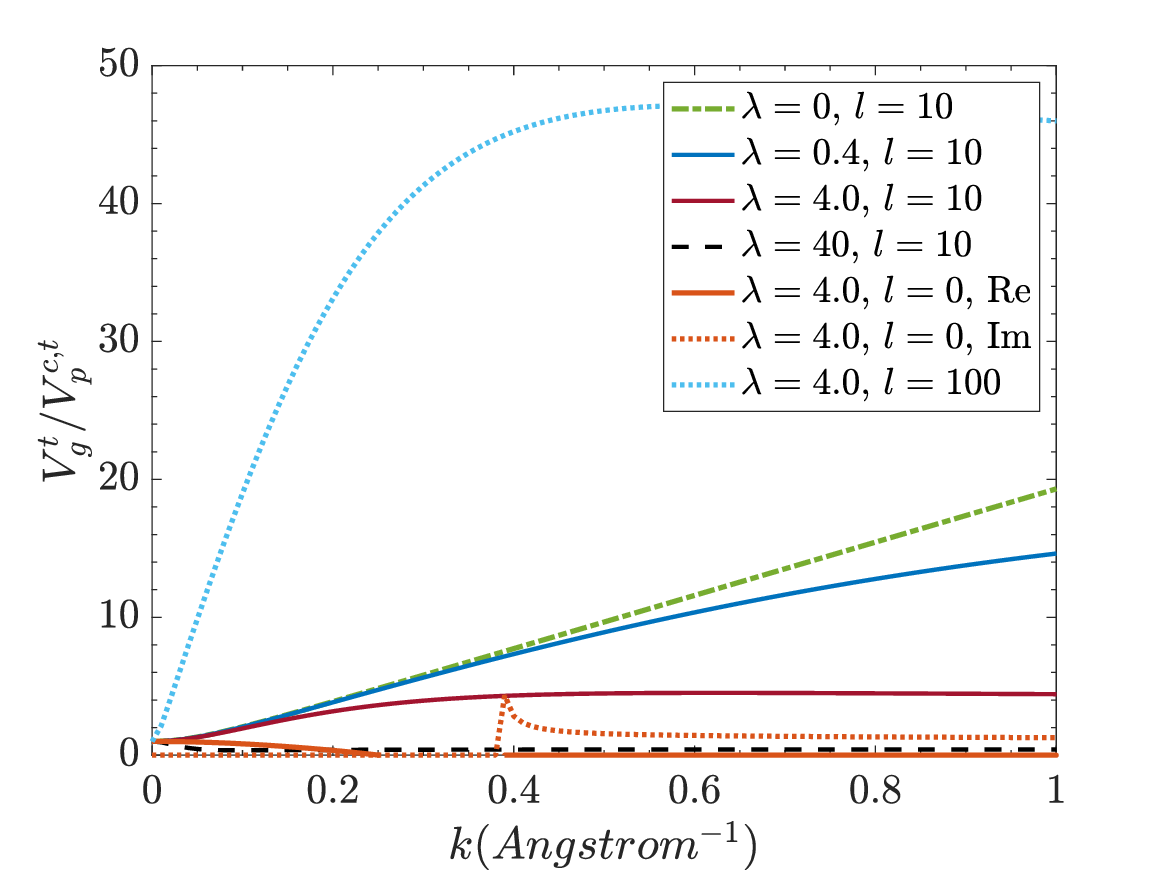}}
{\caption{The effect of the length parameters $\lambda$ and $l$ on the group velocities of (a)-(b) longitudinal and (c)-(d) transverse waves is shown for different values of wave number $k$. The group velocities are normalized by the classical phase velocities. (b) and (d) shows an inset of (a) and (c), respectively, from wave numbers 0 to 1 Angstrom$^{-1}$, denoted by the black box in (a) and (c).}\label{Fig:VgVpc}}
\end{figure}

\paragraph{Group velocity:}
Figure \ref{Fig:VgVpc} shows the influence of the length parameters $\lambda$ and $l$ on the normalized group velocities of longitudinal and transverse waves. The group velocity when $\lambda$ =0 and $l$ = 10 Angstrom is unbounded and non-physical. As $\lambda$ is increased to $0.4$, $4.0$, and $40$ Angstroms, the group velocity decreases and becomes bounded. When the nonlocal elastic interaction length $l$ is set to zero, and $\lambda$ = 4.0 Angstrom, the group velocity is complex, and the real part decays to zero beyond $k \approx 0.2$ Angstrom$^{-1}$, and the imaginary part is bounded. When $\lambda$ = 4.0 Angstrom and $l$ = 100 Angstrom, the group velocity increases and is bounded. Overall, the microstructural and nonlocal elastic interaction length-scale parameters have contrasting effects on the group velocity. While increasing $\lambda$ decreases the group velocity, increasing $l$ increases the group velocity.

\begin{figure}[h]\centering
\subfigure[longitudinal]{\includegraphics[keepaspectratio=true,width=0.45\textwidth]{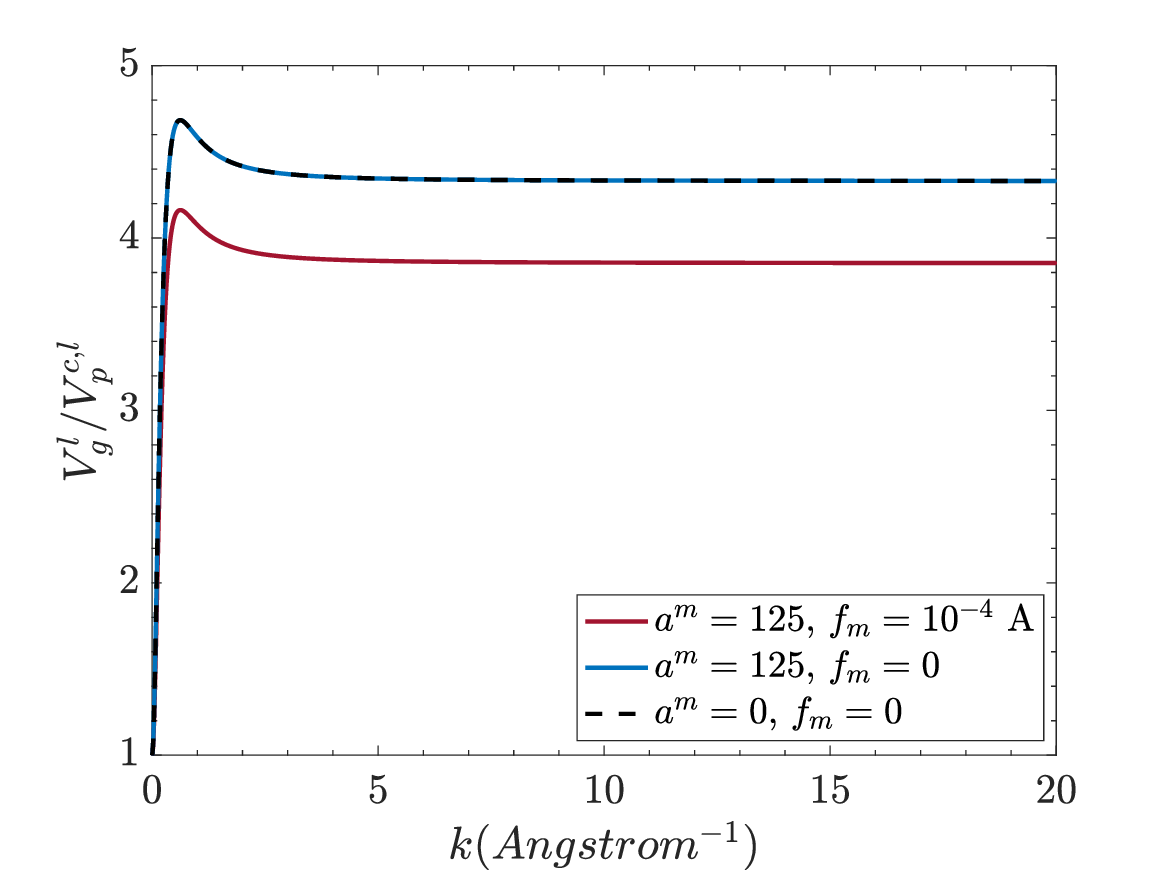}}
\subfigure[transverse]{\includegraphics[keepaspectratio=true,width=0.45\textwidth]{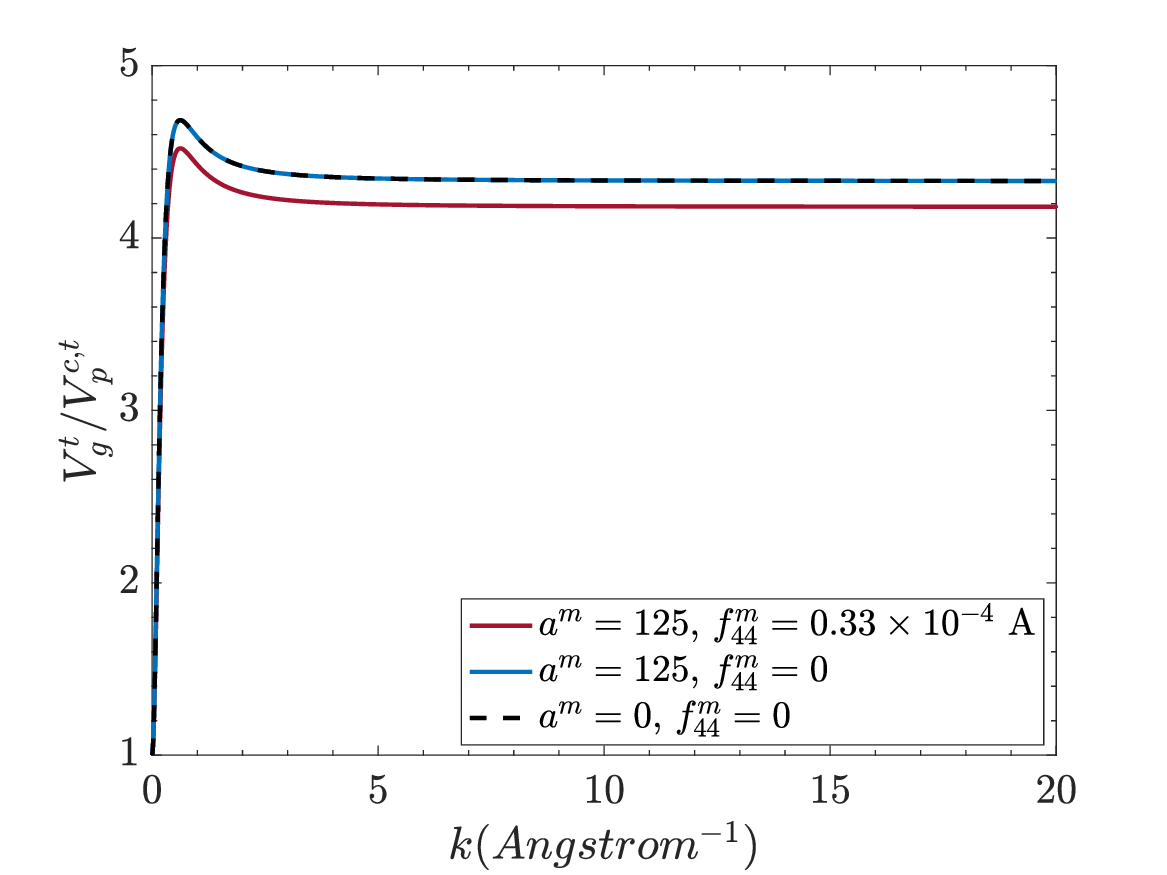}}
{\caption{The effect of magnetism on the group velocities of (a) longitudinal and (b) transverse waves is shown for different values of wave number $k$. The phase velocities are normalized by the classical phase velocities.}\label{Fig:VgVpcMag}}
\end{figure}

In Figure \ref{Fig:VgVpcMag}, we show the effect of magnetism on the group velocities of longitudinal and transverse waves. The values of $\lambda$ and $l$ chosen are 4.0 Angstrom and 10 Angstrom, respectively. Figure \ref{Fig:VgVpcMag} shows that the group velocity decreases when flexomagnetic effects are considered. The group velocity in a non-magnetic material is the same as that in a magnetic but non-flexomagnetic material. This can also be seen by setting $f^m$ = 0 and $f_{12}^m$ = 0 in Equations \ref{Eq:Vgl} and \ref{Eq:Vgt}, respectively. Furthermore, as $f_{m} > f_{44}^m$  and the phase velocity of longitudinal waves depends on $f_{m}$. In contrast, the group velocity of transverse waves depends on $f_{44}^m$. Therefore, flexomagnetism has a stronger influence on the group velocity of longitudinal waves than on that of transverse waves.


\begin{figure}[h] \centering
\subfigure[longitudinal]{\includegraphics[keepaspectratio=true,width=0.45\textwidth]{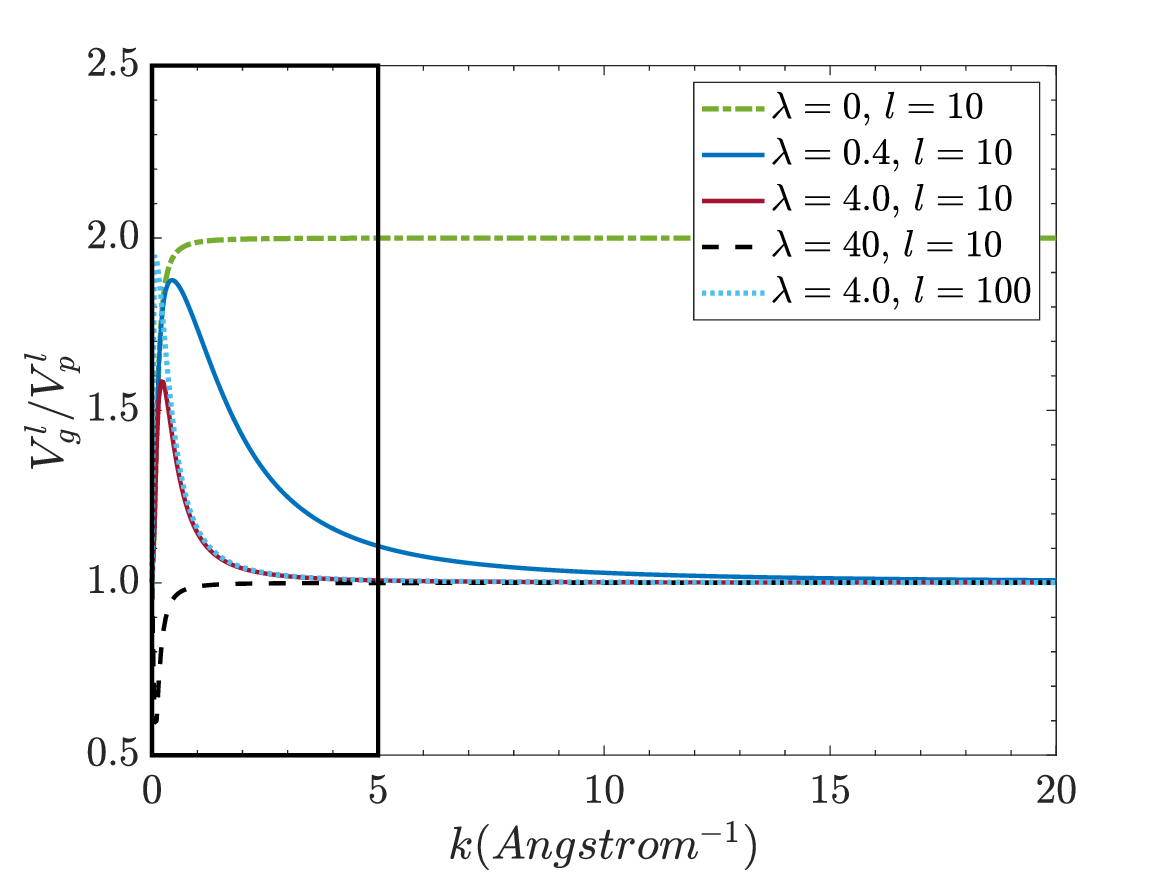}}
\subfigure[longitudinal]{\includegraphics[keepaspectratio=true,width=0.45\textwidth]{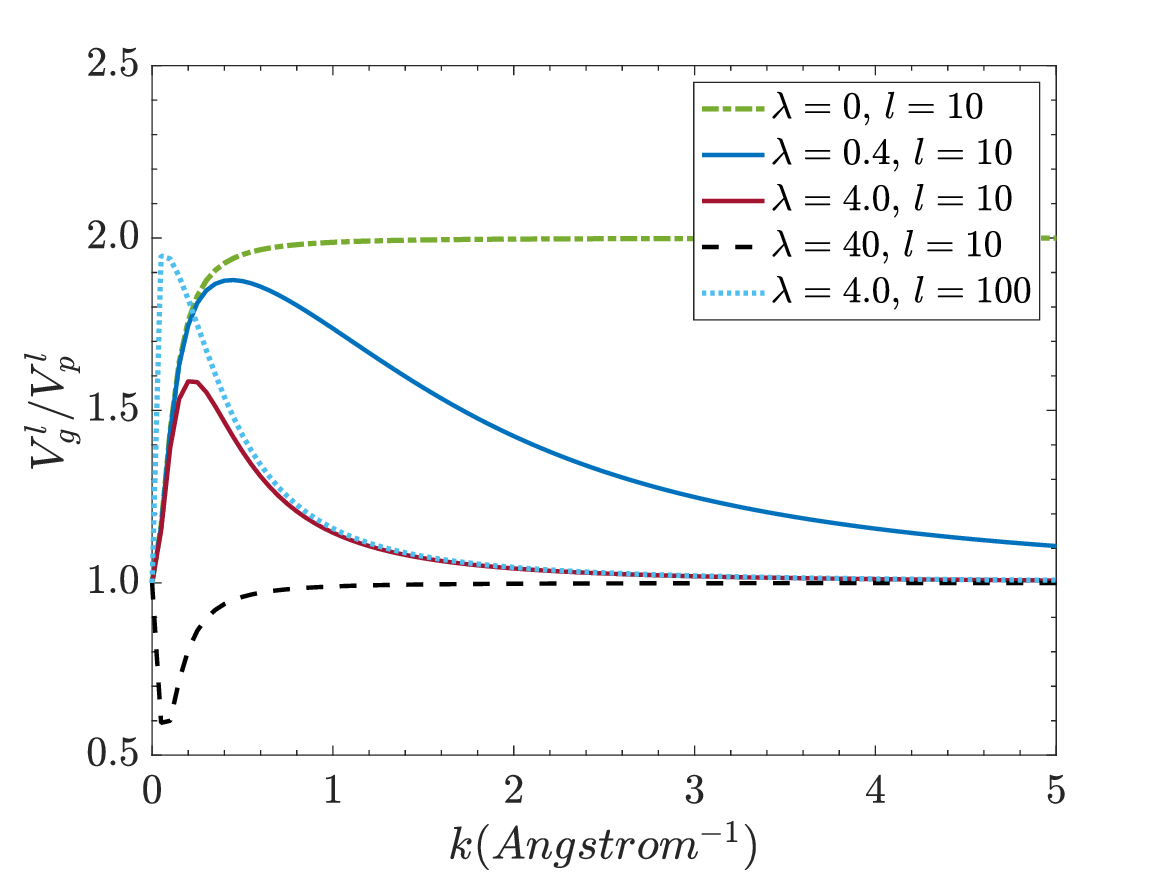}}\\
\subfigure[transverse]{\includegraphics[keepaspectratio=true,width=0.45\textwidth]{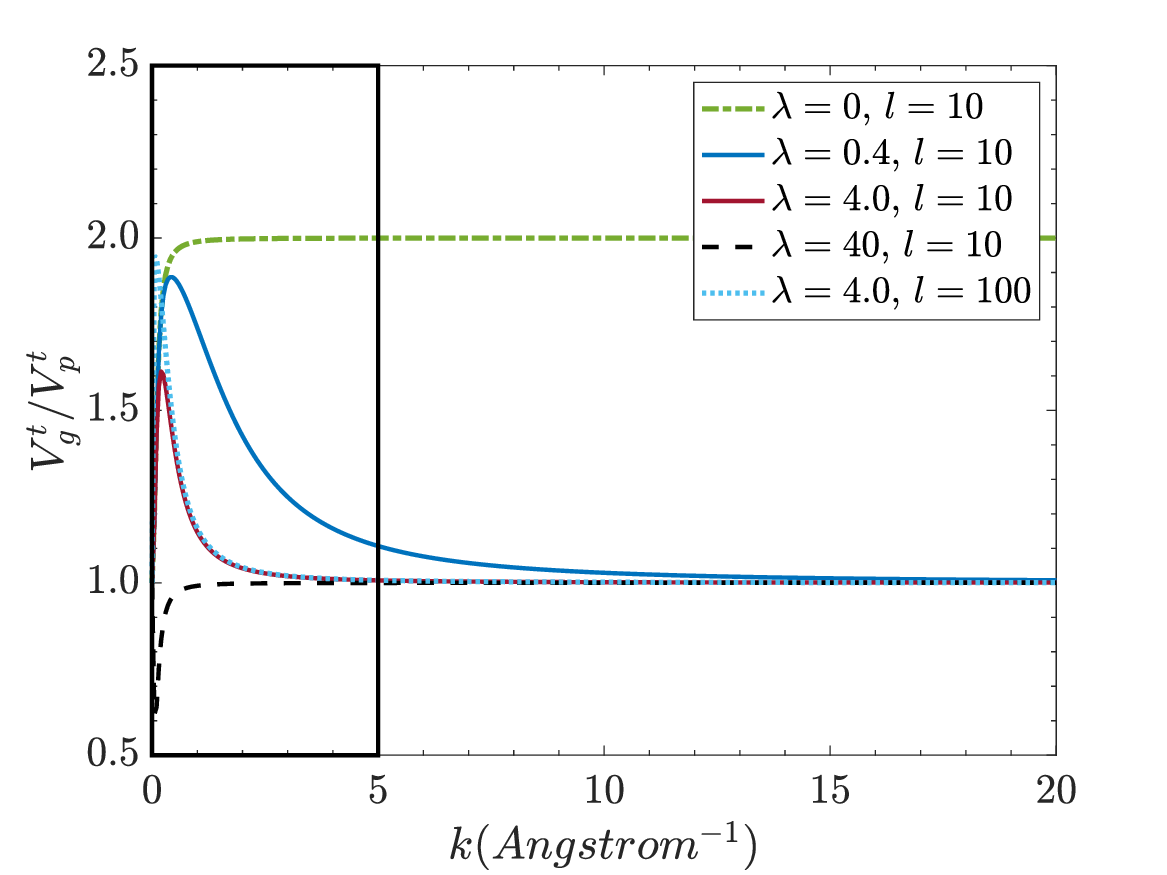}}
\subfigure[transverse]{\includegraphics[keepaspectratio=true,width=0.45\textwidth]{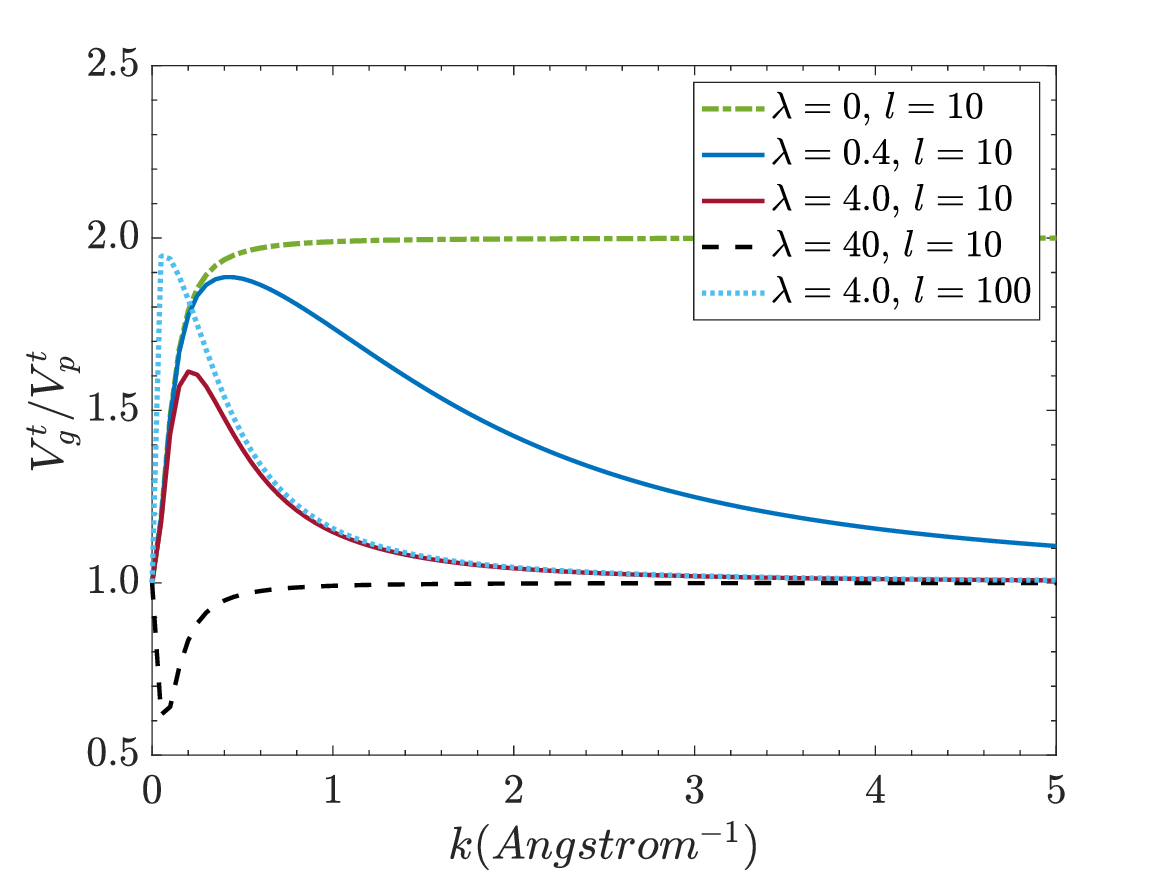}}
{\caption{The effect of the length parameters $\lambda$ and $l$ on the ratio of the group velocity and phase velocity of (a)-(b) longitudinal and (c)-(d) transverse waves are shown for different values of wave number $k$. (b) and (d) shows an inset of (a) and (c), respectively, from wave numbers 0 to 5 Angstrom$^{-1}$, denoted by the black box in (a) and (c).}\label{Ratio}}
\end{figure}

\paragraph{Ratio of group velocity to phase velocity:}
Next, we discuss the ratio of the group velocity to the phase velocity. When this ratio is less than one, the wave shows normal dispersion. When the ratio is greater than one, the wave shows abnormal dispersion, and the wave is non-dispersive if the ratio is one. In classical linear elasticity, the group velocity equals the phase velocity, and wave propagation is non-dispersive. We find that this does not hold for waves traveling in flexomagnetic solids with microstructure and strain-gradient effects. Figure \ref{Ratio} shows the effect of length-scale parameters on the phase velocity-group velocity ratio for longitudinal and transverse waves. At wavenumber $k$=0, the ratio is 1. When the wavenumber is increased, the ratio of phase and group velocity first can initially demonstrates normal or abnormal dispersion, finally decaying to a constant value which depends on $\lambda$ and $l$. When $\lambda$ = 0 and $l$=10 Angstrom, the phase velocity is almost twice the group velocity for wavenumbers greater than 1 Angstrom$^{-1}$, showing abnormal dispersion. Keeping $l$=10 Angstrom, when the microstructural length-scale $\lambda$ is increased to 0.4 and 4.0 Angstrom, the waves first display abnormal dispersion for small wavenumbers, finally decaying to non-dispersive character with increasing wavenumbers. When $\lambda$ is further increased to 40 Angstrom, the waves show normal dispersion for small wavenumbers and transition to non-dispersive waves with increasing wavenumbers. Furthermore, the transition from oscillating between abnormal or normal dispersion to non-dispersive character is quicker with increasing the microstructural length-scale $\lambda$. We also observe that increasing the nonlocal elastic interaction length $l$ from 10 to 100 Angstrom, the waves remain abnormally dispersive for small wavenumbers and non-dispersive for wavenumbers greater than 3 Angstrom$^{-1}$. 

\begin{figure}[h]\centering
\subfigure[longitudinal]{\includegraphics[keepaspectratio=true,width=0.45\textwidth]{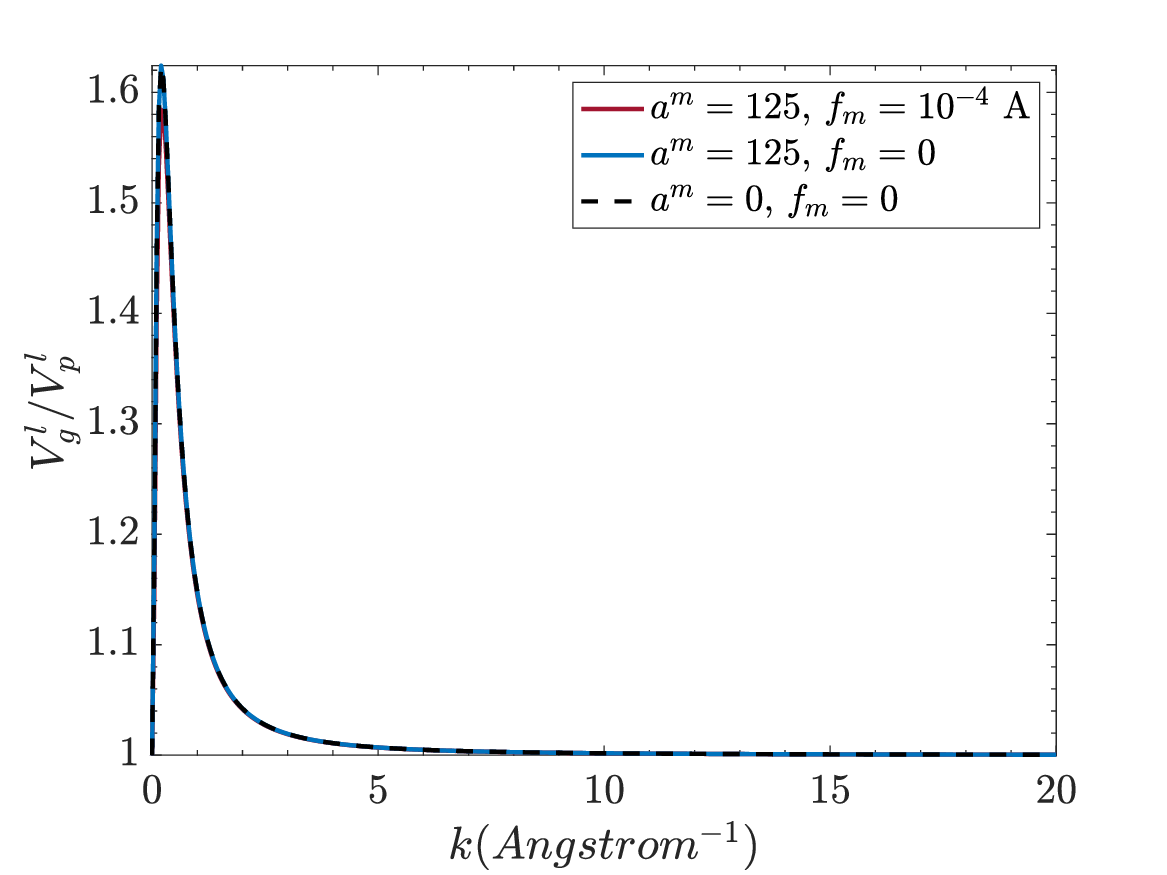}}
\subfigure[transverse]{\includegraphics[keepaspectratio=true,width=0.45\textwidth]{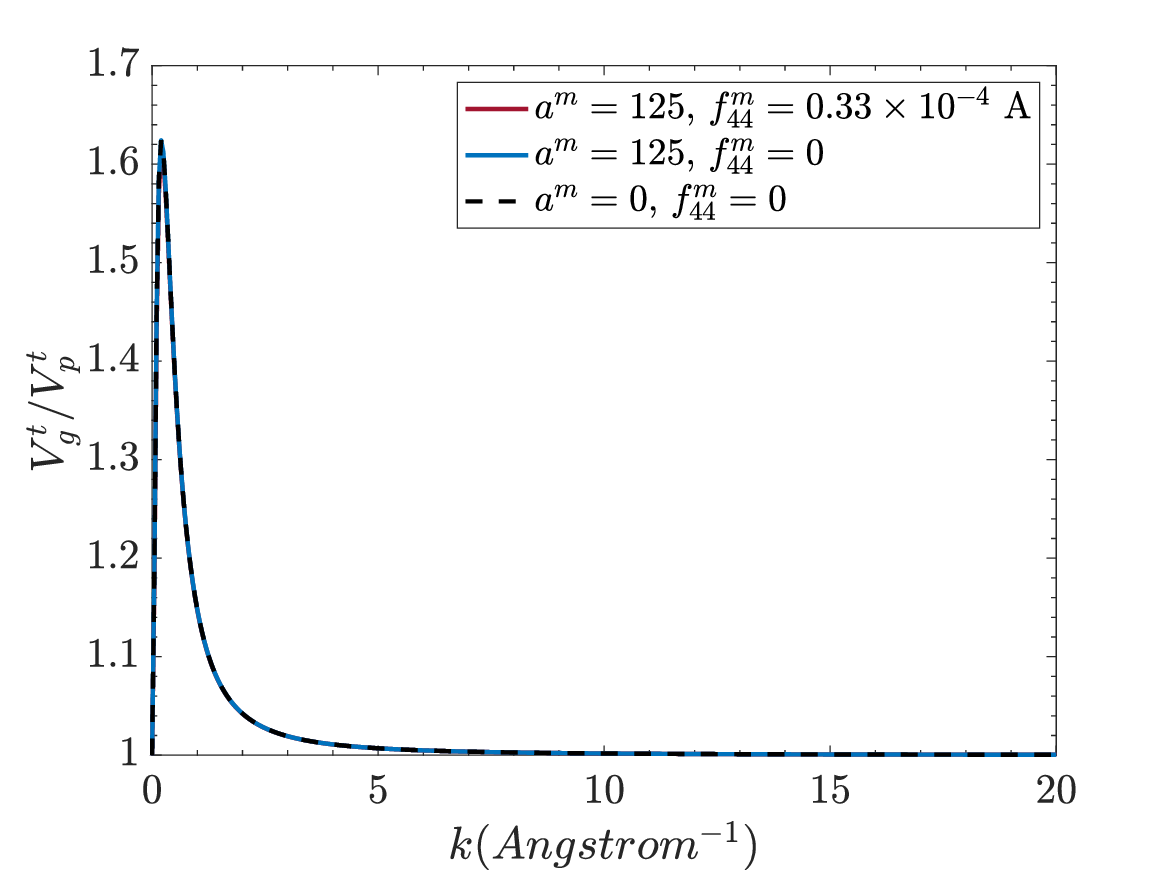}}
{\caption{The effect of magnetism on the ratio of group velocity to phase velocity of (a) longitudinal and (b) transverse waves is shown for different values of wave number $k$. The phase velocities are normalized by the classical phase velocities.}\label{Fig:VgVpMag}}
\end{figure}

 Figure \ref{Fig:VgVpMag} shows the group velocity to the phase velocity ratio for a non-magnetic, magnetic but non-flexomagnetic, and flexomagnetic material. We take $\lambda$ = 4.0 Angstrom and $l$ = 10 Angstrom for this comparison. We see that the ratio of group velocity to the phase velocity is unaffected by magnetism or flexomagnetism.

In essence, the dispersive character of longitudinal and transverse waves can be carefully engineered by tuning the microstructural and nonlocal elastic interaction lengths. In particular, the absence of microstructural length gives rise to abnormally dispersive waves, whereas microstructure ensures the waves are non-dispersive for large wavenumbers and abnormally dispersive for small wavenumbers.

\begin{figure}[h!]\centering
    \subfigure[]{\includegraphics[width=0.45\linewidth]{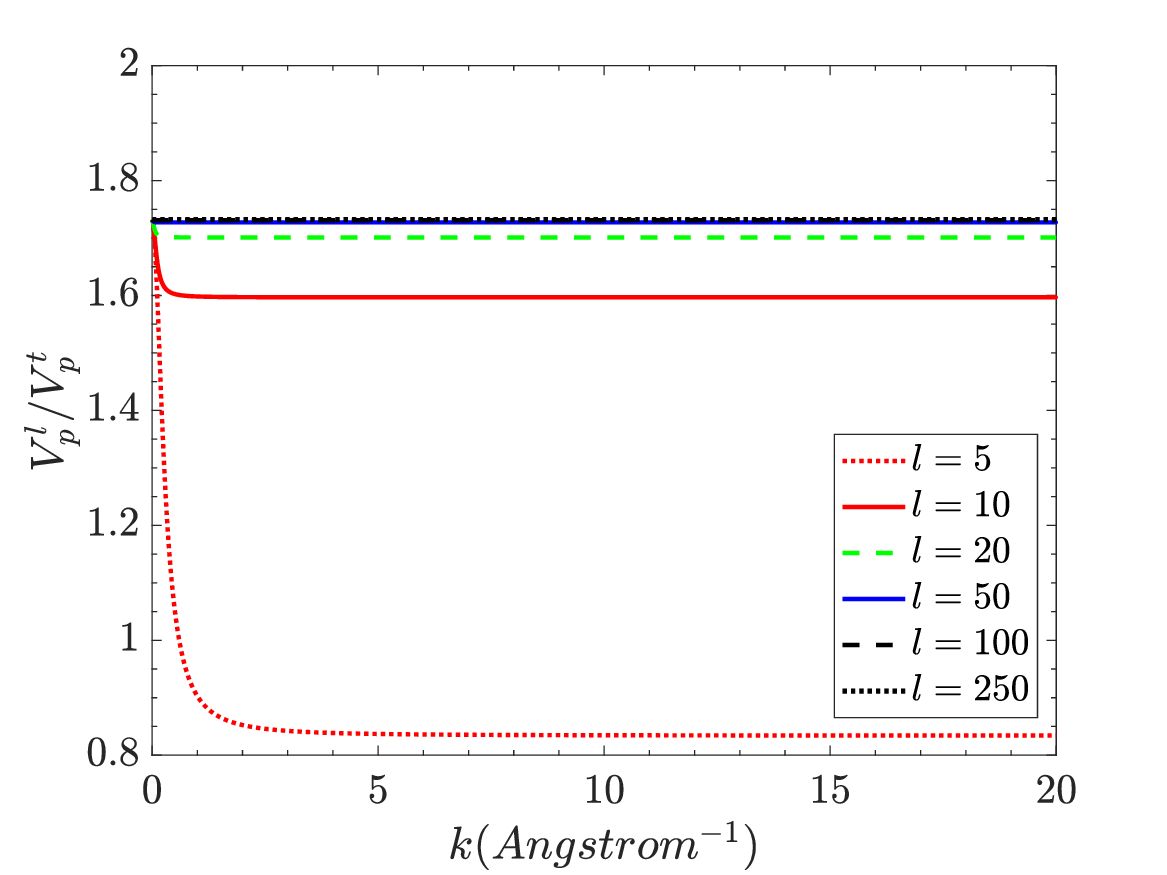}\label{Fig:VplVptl}}
    \subfigure[]{\includegraphics[width=0.45\linewidth]{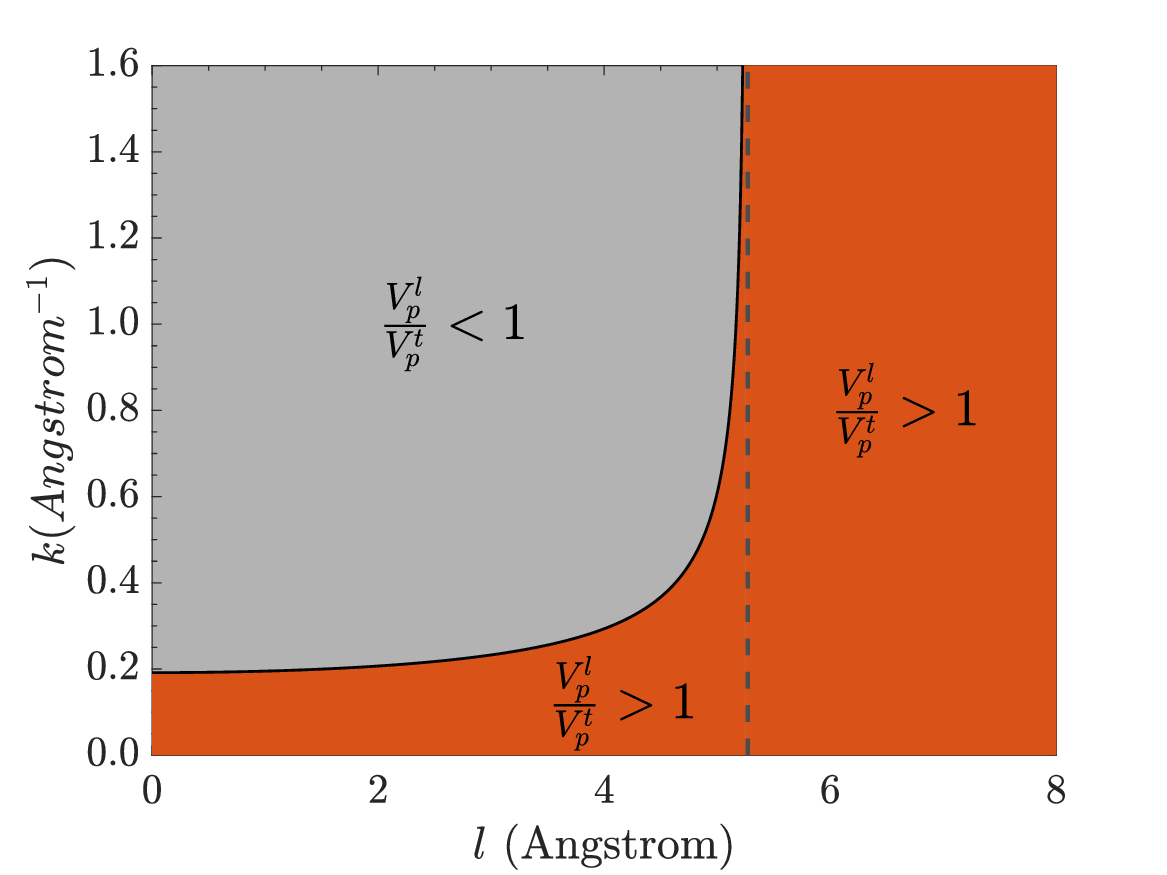}\label{Fig:kstar}}
    \caption{The effect of the nonlocal elastic interaction length $l$ on the ratio of the phase velocities of longitudinal and transverse waves is shown in (a). In (b), the phase diagram shows the region in $k$-$l$ space. The region depicted by the color gray is where phase velocities of transverse waves are more than that of longitudinal waves. The orange region shows the values of $k$ and $l$ for which the phase velocities of longitudinal waves are greater than the phase velocities of transverse waves. The dashed line in (b) is the critical length $l_{crit}$ = 5.27 Angstrom beyond which the phase velocities of longitudinal waves are greater than the phase velocities of transverse waves for all wavenumbers $k$.}
    \label{fig:VplVpt}
\end{figure}

\paragraph{Ratio of phase velocities of longitudinal and transverse waves:}
Finally, we discuss the ratio of the phase velocities of the longitudinal and transverse waves. In an isotropic elastic medium, the ratio of the phase velocities of longitudinal and transverse waves is constant and is given by $\sqrt{\frac{c}{c_{44}}} > 1$. Unlike the classical case, the ratio of longitudinal and transverse phase velocities in an isotropic elastic flexomagnetic solid depends on wavenumber $k$, and is given by
\begin{equation}\label{Eq:LTratio}
    \frac{V_p^{l}}{V_p^{t}} = \left( \frac{c(a_m+1)(k^2l^2 + 1) - \mu_0(k f_{m})^{2}}{c_{44}(a_m+1)(k^2l^2 + 1) - \mu_0(k f_{44}^{m})^{2}}\right)^{\frac{1}{2}} \,\,.
\end{equation}
Figure \ref{Fig:VplVptl} shows the effect of the nonlocal elastic interaction lengthscale $l$ on the ratio of the phase velocities of longitudinal and transverse waves. We vary $l$ from 5 Angstrom to 250 Angstrom. From this plot, we see that at $k$ = 0, the ratio is greater than unity and is independent of $l$. For $l$ = 5 Angstroms, as the wavenumber is increased, the ratio of phase velocities of longitudinal and transverse waves quickly decreases to a value less than unity. The ratio is more than unity when $l$ = 10, 20, 100, and 250 Angstroms. This motivates us to investigate the values of $k$ and $l$ for which phase velocities of longitudinal waves are greater than the phase velocities of transverse waves in an isotropic elastic flexomagnetic medium.

From Equation \ref{Eq:LTratio}, we see that the ratio of phase velocities of longitudinal and transverse waves is greater than unity for
\begin{equation}
   |k| < \,\, k_{crit} \coloneqq \,\,\left( \frac{(c_{12}+c_{44})(a_m+1)}{\mu_0\left[f_m^2 - (f_{44}^m)^2\right] -(c_{12}+c_{44})(a_m+1)l^2}\right)^{\frac{1}{2}} \,\, .
\end{equation}
Notice that $k_{crit}$, is independent of the microstructural length $\lambda$ but depends on the nonlocal elastic interaction length $l$ and $l^2 < l_{crit}^2 \coloneqq \mu_0\left(\frac{f_m^2 - (f_{44}^m)^2}{(c_{12}+c_{44})(a_m+1)}\right)$. When $l>l_{crit}$, then the ratio of longitudinal and transverse phase velocities is greater than unity for all wavenumbers. Using the values in Table \ref{parametertable}, we obtain $l_{crit}$ = 5.27 Angstrom. Also, in the absence of flexomagnetism, the critical length is zero and thus ratio of longitudinal and transverse phase velocities is greater than unity.

In Figure \ref{Fig:kstar}, we plot the regions in the $k$-$l$ parameter space where the orange region shows the values of $k$ and $l$ for which phase velocities of longitudinal waves are greater than the phase velocities of transverse waves, and the gray region shows the value of $k$ and $l$ for which transverse waves travel faster than longitudinal waves, which is not observed in a material with classical elasticity. 

\subsection{Attenuation of elastic waves in a flexomagnetic solid}\label{Sec:Attenuation}
In this section, we discuss conditions under which elastic waves can attenuate in a flexomagnetic solid. For propagating waves, also known as traveling waves, the wave numbers $k$ are real. However, when the wavenumber is imaginary, the wave is attenuated and is known as a standing wave or evanescent wave, characterized by a zero cutoff frequency. The phenomenon of attenuation of mechanical waves in a flexoelectric material has been studied by \cite{Giannakopoulos2024}. We follow their analysis for waves in flexomagnetic material. To encompass both longitudinal and transverse cases, we write the elastic modulus as
\begin{eqnarray}\label{cbar}
   \Bar{c} = \begin{cases} 
   c\,,\,\,\,\, \text{longitudinal waves}\\
    c_{44} \,, \text{transverse waves} \,
\end{cases}
\end{eqnarray}
and the flexomagnetic constant as
\begin{eqnarray}\label{fbar}
   \Bar{f_m} = \begin{cases} 
   f_m\,,\, \text{longitudinal waves}\\
    f_{44}^m \,, \text{transverse waves} \,
\end{cases}
\end{eqnarray}

We solve Equation \ref{Eq:omegal} or \ref{Eq:omegat} for the wave number $k$ as a function of wave frequency $\omega$ and obtain two roots $k_1$ and $k_2$
\begin{equation}\label{Eq:k1square}
    k_1^2(\omega) = \frac{2 \omega^2 \rho}{\Bar{c}-\frac{\lambda^2}{3} \omega^2 \rho  + \sqrt{\left(\Bar{c}-\frac{\lambda^2}{3} \omega^2 \rho \right)^2 + 4 \left( \Bar{c} l^2 - \frac{\mu_0\Bar{f_m}^2}{1+a_m}\right)^2 \omega^2 \rho }} \geq 0 \,\,,
\end{equation}
and
\begin{equation}\label{Eq:k2square}
    k_2^2(\omega) = \frac{2 \omega^2 \rho}{\Bar{c}-\frac{\lambda^2}{3} \omega^2 \rho  - \sqrt{\left(\Bar{c}-\frac{\lambda^2}{3} \omega^2 \rho \right)^2 + 4 \left( \Bar{c} l^2 - \frac{\mu_0\Bar{f_m}^2}{1+a_m}\right)^2 \omega^2 \rho }} \leq 0 \,\,.
\end{equation}

We limit our analysis to real wave frequencies. From the above two expressions it is easy to see that $k_1$ is always real and is associated with propagating waves or traveling waves. The wave number $k_2$ is always imaginary and is associated with standing waves or evanescent waves. This can be seen by writing $k_2=i|k_2|$, and substituting it in the ansatz for displacement 
\begin{eqnarray}
    u(x,t) &=& u_0 \exp[i(k_2 x-\omega t)] \,\,, \nonumber \\
        &=& u_0 \exp(-|k_2|x)\, \exp(-i \omega t) \,\,,
\end{eqnarray}
where the first term in the last expression leads to exponential decay of the displacement field in space, leading to attenuation of elastic waves. From Expressions \ref{Eq:k1square} and \ref{Eq:k2square}, it is easy to see that $|k_1|<|k_2|$ and $|k_1| \leq \left[ \frac{\bar{c}(a_m+1)}{\mu_0 \bar{f_m}^2 - \bar{c}(a_m+1)l^2}\right]^{1/2}$. 

For attenuated waves, the wave frequency is
\begin{equation}\label{Eq:omegak2}
    {\omega}^2 = |k_2|^2 \left( \frac{\bar{c}(a_m+1)(|k_2|^2l^2-1) - \mu_0 |k_2|^2 \bar{f_m}^2}{\rho (a_m+1) (1-\frac{1}{3}|k_2|^2 \lambda^2)} \right)
\end{equation}
In (\ref{Eq:omegak2}), the limits $\omega \to 0$ yields $|k_2| = \sqrt{3}/{\lambda} $ and $\omega \to \infty$ yields $|k_2| = \left({l^2 - \frac{\mu_0 \bar{f_m}^2}{\bar{c}(a_m+1)} }\right)^{-1/2}$. Therefore, we can write the following range of values of $|k_2|$
\begin{equation}
    \frac{\sqrt{3}}{\lambda} < |k_2| < {\left({l^2 - \frac{\mu_0 \bar{f_m}^2}{\bar{c}(a_m+1)} }\right)^{-1/2}} \,\,.
\end{equation}
In the absence of flexomagnetism, $\bar{f_m}=0$ we obtain $\frac{\sqrt{3}}{\lambda} < |k_2| < \frac{1}{l}$ and . 

Notice that when $l=\lambda/\sqrt{3}$, and $f_m=0$, then $k_1 = \omega \sqrt{\rho/c}$, $k_2 = i/l$, and when $l=\lambda=f_m=0$, then $k_1 = \omega \sqrt{\rho/c}$ and $k_2 \to \infty$. Therefore, elastic waves can be attenuated even in the absence of flexomagnetic effects due to material or microstructural features.

We remark that attenuated waves are not observed in classical elastodynamics (\cite{Giannakopoulos2024,Achenbach}). However, such waves are observed in flexoelectric material (\cite{Giannakopoulos2024}), at free surfaces of anti-plane couple stress materials (\cite{Nobili2020}), and biomaterial surface modeled using second gradient elastic continua (\cite{dellisola2012}). 

\subsection{Zero and negative group velocity of propagating waves}\label{Sec:ZeroandNegativeGV}

The group velocity quantifies energy transport and the evolution of the wave packet. Anomalous features in the dispersion relation, such as zero and negative group velocities, are critical regimes of wave propagation that exhibit interesting properties. Zero group velocity modes arise at the stationary points of the frequency-wavenumber spectrum ($\partial \omega / \partial k = 0$), resulting in the formation of non-propagating standing waves and strong local energy confinement (\cite{prada2008local,prada2009influence,kausel2012number}). This phenomenon is characterized in the higher-order modes of elastic waveguides (\cite{mindlin1951influence,clorennec2007local}), and in periodic media exhibiting "frozen mode" degeneracies( \cite{figotin2006frozen,Fishman2024,Deshmukh2024}). In Section \ref{Sec:WaveFreezing} we present a detailed discussion on the phenomenon of wave freezing. Conversely, in the negative group velocity regime, the wavepackets move in the opposite direction (backwards) to the wave propagation direction (forward). The investigation of the negative group velocity phenomenon has attracted attention (\cite{meitzler1965,dolling2006simultaneous,martinez1984negative,ye2013negative}). Specifically, locally resonant microstructures or inertial amplification mechanisms can induce effective constitutive properties that support backward energy transport (\cite{liu2000locally,willis2011effective}). These contemporaneous advances in the study of zero- and negative-group-velocity modes in materials motivate us to investigate these phenomena for flexomagnetic solids.

For zero and negative group velocity,  setting $V_g \leq 0$ in (\ref{Eq:Vgl}) for longitudinal waves or (\ref{Eq:Vgt}) for transverse waves, and using (\ref{cbar}) and (\ref{fbar}), we obtain the following condition 
\begin{equation}\label{Cond1}
    \bar{c}(a_m+1)(k^2 l^2 + 1) - k^2 \mu_0 \bar{f_m}^2 \neq 0  \,\,,
\end{equation}
and
\begin{equation}\label{Cond2}
   \left(\bar{c}(a_m+1)l^2 - \mu_0 \bar{f_m}^2 \right)(\lambda^2 k^4 + 6k^2) +3 \bar{c} (a_m+1) \leq 0 \,\,.
\end{equation}
For convenience, we denote
\begin{eqnarray}
   && \beta \coloneqq \mu_0\bar{f_m}^2-\bar{c}(a_m+1)l^2 \,\,, \label{Eq:Beta} \\
   && \gamma \coloneqq 3\bar{c}(a_m+1) \,\,.\label{Eq:Gamma}
\end{eqnarray}

Condition (\ref{Cond1}) yields
\begin{equation}\label{kneq}
    |k| \neq \left[ \frac{\bar{c}(a_m +1)}{\mu_0 \bar{f_m}^2 - \bar{c}(a_m +1)l^2} \right]^{1/2} = \sqrt{\frac{\gamma}{3\beta}} \,\,,
\end{equation}
and $\frac{\gamma}{\beta} > 0$, $\beta \neq 0$. If $\frac{\gamma}{\beta} < 0$, then condition (\ref{Cond1}) is always satisfied for all real values of $k$.

To evaluate the second condition in inequality (\ref{Cond2}), two cases are considered. First with $\lambda = 0$, and second with $\lambda \neq 0$.

\paragraph{Case 1: Microstructure is absent, $\lambda = 0$}
In this case, the inequality (\ref{Cond2}) is 
\begin{equation}\label{Cond2b}
   - 6 \beta k^2 + \gamma \leq 0 \,\,.
\end{equation}
Solving above, zero group velocity is obtained for wavenumber
\begin{equation}\label{Eq:ZeroGVRangeC1}
|k| = \sqrt{\frac{\gamma}{6\beta}} \,\,.
\end{equation}
From (\ref{Cond2b}) and (\ref{kneq}) the range of $k$ for which the group velocity is negative is 
\begin{equation}\label{Eq:NegativeGVRangeC1}
    |k| \in \left\{\left( \sqrt{\frac{\gamma}{6\beta}}\,,\, \infty\right) \;\middle|\; |k| \neq  \sqrt{\frac{\gamma}{3\beta}}\,,\, \frac{\gamma}{\beta} > 0 \,,\, \beta \neq 0 \right\} \,\,.
\end{equation}
The condition $\gamma/\beta > 0 $ yields $l^2 < \frac{\mu_0 \bar{f_m}^2}{\bar{c}(a_m+1)}$. When $\gamma/\beta < 0$, and correspondingly $l^2 > \frac{\mu_0 \bar{f_m}^2}{\bar{c}(a_m+1)}$, then the waves are attenuated and has been discussed Section \ref{Sec:Attenuation}.  Our analysis also show that when $l^2 = \frac{\mu_0 \bar{f_m}^2}{\bar{c}(a_m+1)}$, then $\beta=0$, and, we have a singularity. 

We remark that in the absence of flexomagnetism i.e. $\bar{f_m} = 0$, we obtain $\gamma/\beta = -3/l^2 < 0$, and therefore the group velocity of propagating waves are strictly positive in such a scenario. When flexomagnetic effects are considered, for the values of material constants listed in Table \ref{parametertable}, $\gamma/\beta < 0$ for $l > 4.5583$ Angstrom for longitudinal waves and $l > 2.6054$ Angstrom for transverse waves. Thus, for a typical choice $l=10$ Angstrom, zero and negative group velocities are not observed. However, when non-local elastic interaction is absent (i.e. $l=0$), but flexomagnetic effects are present, then $\gamma/\beta>0$, therefore, zero and negative group velocity is observed and we estimate $\sqrt{\frac{\gamma}{6\beta}}$ in Equations \ref{Eq:ZeroGVRangeC1} and \ref{Eq:NegativeGVRangeC1} as $0.1551$ Angstrom$^{-1}$ for longitudinal and $0.2714$ Angstrom$^{-1}$ for transverse waves.

\paragraph{Case 2: Microstructure is present, $\lambda > 0$} 
In this case, the inequality \ref{Cond2} is 
\begin{equation}
  -\lambda^2 \beta k^4 - 6 \beta k^2 + \gamma \leq 0 \,\,.
\end{equation}
Solving the above, we obtain two wavenumbers for zero group velocity
\begin{equation}\label{Eq:ZGVk1}
 k_1^{2} = \frac{3-\sqrt{3+\lambda^2 (\gamma/\beta)}}{-\lambda^2} \geq 0\,\,.
\end{equation}
and
\begin{equation}\label{Eq:ZGVk2}
 k_2^{2} = \frac{3+\sqrt{3+\lambda^2 (\gamma/\beta)}}{-\lambda^2} \leq 0\,\,.
\end{equation}
From above, we see that $k_2$ is imaginary and hence the waves propagating with wavenumber $k_2$ will be attenuated.

We first analyze the case when $\gamma/\beta > 0$ and $\beta \neq 0$.  The range of wavenumbers for which the group velocity will be negative is obtained from $k_1$ and is
\begin{equation}\label{Eq:NegativeGVRange}
    |k| \in \left\{\left(  \left [\frac{\sqrt{3+\lambda^2 (\gamma/\beta)}-3}{\lambda^2} \right]^{1/2} \,,\, \infty\right) \;\middle|\; |k| \neq  \sqrt{\frac{\gamma}{3\beta}}\,,\, \frac{\gamma}{\beta} > 0 \,,\, \beta \neq 0 \,, \lambda^2 \geq \frac{6\beta}{\gamma} \right\} \,\,,
\end{equation}
and the group velocity will be zero for the wavenumber 
\begin{eqnarray}\label{Eq:ZeroGVRange}
    |k| = \left [\frac{\sqrt{3+\lambda^2 (\gamma/\beta)}-3}{\lambda^2} \right]^{1/2} \,\, \text{when} \,\, \lambda^2 \geq \frac{6\beta}{\gamma} .
\end{eqnarray}
 Using the relation $\lambda^2\geq \frac{6\beta}{\gamma}$, we obtain the following relation for $\lambda$ and $l$
\begin{eqnarray}\label{llambdarelation}
    l^2 + \frac{\lambda^2}{2} \geq  \frac{\mu_0 \bar{f_m}^2}{\bar{c}(a_m+1)}\,\,.
\end{eqnarray}

Next, we analyze the case when $\gamma/\beta < 0$ and $\beta \neq 0$. In this case, $k_1$ is always imaginary, and zero or negative group velocities are not observed for real wavenumbers. In particular, in the absence of flexomagnetism (i.e. $\bar{f}_m=0$ ), we have $\gamma/\beta = -3/l^2 <0$. 

In the presence of flexomagnetism,  when $l^2 < \frac{\mu_0 \bar{f_m}^2}{\bar{c}(a_m+1)}$ then $\gamma/\beta>0$. Using the parameters listed in Table \ref{parametertable}, we obtain $l<4.5583$ Angstrom and $l< 2.605$ Angstrom for longitudinal and transverse waves, respectively, and the wavenumbers for negative and zero group velocity are given by (\ref{Eq:NegativeGVRange}) and (\ref{Eq:ZeroGVRange}), respectively. When $l^2 > \frac{\mu_0 \bar{f_m}^2}{\bar{c}(a_m+1)}$, then $\gamma/\beta<0$, and zero or negative group velocity are not observed. In particular, when non-local elastic interactions are absent (i.e. $l=0$), we have $\gamma/\beta>0$. In this case, using \ref{llambdarelation}, we see that when $\lambda \geq 6.446$ Angstrom for longitudinal waves or $\lambda \geq 3.684$ Angstrom for transverse waves, then zero and negative group velocity are observed for wavenumbers (\ref{Eq:ZeroGVRange}) and (\ref{Eq:NegativeGVRange}), respectively. For the typical choice of parameters $l=10$ and $\lambda=4$ Angstrom, we have $\gamma/\beta<0$, and negative or zero group velocity modes are absent.

\subsection{Wave freezing in a flexomagnetic solid}\label{Sec:WaveFreezing}
\emph{Wave freezing} is the phenomenon where a propagating wave stops at a location in space without diffusing or spreading (\cite{figotin2006frozen,Fishman2024,Deshmukh2024}). For such a phenomenon to occur, the group velocity of a propagating wave should be zero, while the phase velocity remains non-zero, and the derivative of the group velocity taken with respect to the wavenumber should be zero i.e. stationary inflection point. This "freezes" the wave energy in a localized spatial domain. It should be emphasized that this form of control and localization of mechanical energy is distinct from bandgap attenuation.  

The phenomenon of wave freezing was initially investigated in optics (\cite{figotin2006frozen}), and has been recently explored in the elastodynamics of metamaterials. For instance, \cite{Fishman2024} demonstrated that third-order exceptional points in planar elastic laminates can induce axially frozen modes, where mode coalescence results in a vanishing axial group velocity despite finite transmittance. Furthermore, wave freezing can be realized in time-varying metamaterials through the adiabatic modulation of effective properties; \cite{Trainiti2019} established that time-periodic stiffness modulation enables non-reciprocal stop bands that can arrest wave propagation. These mechanisms offer transformative potential for applications requiring high-density elastic energy storage, enhanced sensing via prolonged wave-matter interaction, and precise delay lines. In the recent work of \cite{Deshmukh2024}, wave-freezing has been extended to space–time modulated systems and temporal metasurfaces, where tailored time-dependent material properties can halt wave packets without diffusion through zero-group-velocity modes in engineered phononic lattices. These developments motivate us to explore the possibility of observing such phenomena in a linear elastic flexomagnetic material. 

Waves are frozen when the group velocity is zero ($ \frac{\partial \omega}{\partial k}=0$) and the derivative of the group velocity is also zero ($ \frac{\partial^2 \omega}{\partial k^2}=0$) 


From Equations \ref{Eq:omegal} and \ref{Eq:omegat}, the second derivative of wave frequency expressed in terms of $\beta$ (Equation \ref{Eq:Beta}) and $\gamma$ (Equation \ref{Eq:Gamma}) is
\begin{equation}\label{Eq:omegasecondderivative}
     \frac{\partial^2 \omega}{\partial k^2} = \frac{\omega \left( 3\beta +   \frac{\lambda^2 \gamma}{3}\right) \bigg( -\lambda^2 \beta k^4 + 6\beta k^2 - 3\gamma \bigg)}{\bigg(k^2\lambda^2+3 \bigg)^2 \bigg(-\beta k^2 + \frac{\gamma}{3}\bigg)^2} \,\,.
\end{equation}

Looking at the denominator in (\ref{Eq:omegasecondderivative}), and restricting our analysis to real wavenumbers, we can write,
\begin{equation}
    |k| \neq \sqrt{\frac{\gamma}{3\beta}}
\end{equation}
Notice the equality in the above expression also renders the wave frequency $\omega = 0$.

Next, from the numerator of \ref{Eq:omegasecondderivative}, we obtain two equations
\begin{equation}\label{WFcond1}
    3\beta +   \frac{\lambda^2 \gamma}{3} = 0 \,\,,
\end{equation}
or
\begin{equation}\label{WFcond2}
     -\lambda^2 \beta k^4 + 6\beta k^2 - 3\gamma = 0\,\,.
\end{equation}
From the first equation (\ref{WFcond1}), we get
\begin{equation}\label{Eq:lambdacond}
    \lambda^2 = -\frac{9\beta}{\gamma}\,\,.
\end{equation}
From (\ref{Eq:lambdacond}), we obtain the relation 
\begin{eqnarray}\label{llambdarelationWF}
    l^2 - \frac{\lambda^2}{3} = \frac{\mu_0 \bar{f_m}^2}{\bar{c}(a_m+1)} \,\,.
\end{eqnarray}
Since $\lambda$ is real number, then $\beta/\gamma < 0$, and correspondingly $l^2 > \frac{\mu_0 \bar{f_m}^2}{\bar{c}(a_m+1)}$. In this case, following our discussion in section \ref{Sec:ZeroandNegativeGV}, the waves attenuate.  

We evaluate the second equation (\ref{WFcond2}) for two cases. First, in the absence of microstructure ($\lambda=0$), and next when the microstructure is present ($\lambda \neq 0$).

\paragraph{Case 1: Microstructure is absent, $\lambda = 0$}
In this case, we can write (\ref{WFcond2}) as
\begin{equation}
    6 \beta k^2 - 3\gamma = 0 \,\,,
\end{equation}
and solving above, we obtain
\begin{equation}\label{Eq:WFlambdazero}
    |k| = \sqrt{\frac{\gamma}{2\beta}} \,\,.
\end{equation}
As $\sqrt{\frac{\gamma}{2\beta}} > \sqrt{\frac{\gamma}{6\beta}}$, then from section \ref{Sec:ZeroandNegativeGV}, the group velocity for the wavenumber in (\ref{Eq:WFlambdazero}) is negative, and corresponds to zero group velocity dispersion i.e. wave packets propagate without spreading to second order.

\paragraph{Case 2: Microstructure is present, $\lambda > 0$}
In this case, we obtain
\begin{equation}
    -\lambda^2 \beta k^4 + 6\beta k^2 - 3\gamma = 0
\end{equation}
Solving the above equation, we obtain two roots
\begin{equation}\label{Eq:WFk1}
    k_1^2 = \frac{3}{\lambda^2} \left( 1+ \sqrt{1-\left( \frac{\lambda^2 \gamma}{3\beta}\right)}\right) \,\,,
\end{equation}
and
\begin{equation}\label{Eq:WFk2}
    k_2^2 = \frac{3}{\lambda^2} \left( 1- \sqrt{1-\left( \frac{\lambda^2 \gamma}{3\beta}\right)}\right) \,\,.
\end{equation}

We first analyze the case when $\gamma/\beta > 0$ and $\beta \neq 0$. In this case, both $k_1$ and $k_2$, is real when $\lambda^2 \leq {\frac{3\beta}{\gamma}}$.  Equating these wavenumbers to the wavenumbers in (\ref{Eq:NegativeGVRange}) or (\ref{Eq:ZeroGVRange}), we see there are no solutions, and hence wave freezing is not observed for this case.

For $\gamma/\beta < 0 ,\, \beta \neq 0$, we see that $k_1$ is real and $k_2$ is imaginary. From the discussion in section (\ref{Sec:ZeroandNegativeGV}), we know that zero or negative group velocity modes are absent when  $\gamma/\beta < 0$. Therefore, wave freezing is not observed in this case.





\section{Concluding remarks}\label{Sec:conclusion}
In this work, we have derived a theory of propagation of longitudinal and transverse elastic waves in a linear elastic flexomagnetic solid with microstructure and non-local interactions. The effect of microstructure and non-local parameters is accounted for using length-scale parameters. The frequency, phase velocity, and group velocity depend on the flexomagnetic coefficient and the length-scale parameters. In a classical elastic material, propagating waves are non-dispersive. Our analysis shows that, depending on the choice of the length-scale parameters, the waves can exhibit normal or abnormal dispersion. Also, unlike classical elasticity, where the ratio of phase velocities of longitudinal and transverse waves is greater than unity, in a flexomagnetic solid, this ratio can be less than unity for a certain choice of non-local interaction length and wavenumbers. We have also shown that waves in a flexomagnetic solid can attenuate for a range of wavenumbers that depend on the flexomagnetic coefficient and microstructural parameters. Finally, we have discussed the possibility of obtaining zero- and negative-group-velocity modes, as well as the phenomenon of wave freezing in a flexomagnetic solid. 

We conclude by outlining several directions for future investigation that arise from this work. Here, we have assumed material isotropy; however, many nanoscale materials are intrinsically anisotropic (\cite{dresselhaus2025anomalous}), making it important to examine how anisotropy influences wave propagation in flexomagnetic media. In addition, some materials may exhibit flexoelectric effects alongside flexomagnetism, motivating a natural extension of the theory to account for both couplings simultaneously.

Recent efforts have focused on developing nonlinear theories of flexoelectricity (\cite{Deng2014,Grasinger2021,codony:Finite}) and flexomagnetism (\cite{sky2025cosserat}), yet the problem of wave propagation in nonlinear flexoelectric and flexomagnetic materials remains largely unexplored. Another promising direction is dynamic flexomagnetism, in which magnetization dynamics contribute to the kinetic energy of the system and become particularly significant at high wave numbers. This effect may be viewed as analogous to the dynamic flexoelectric effect (\cite{Yudin2013Review}). To the best of our knowledge, dynamic flexomagnetism has not yet been systematically investigated, and experimental and first-principles determination of the dynamic flexomagnetic tensor is essential for incorporating these effects into future theoretical frameworks.

\section*{Acknowledgements}
 This research used resources of the Oak Ridge Leadership Computing Facility, a DOE Office of Science User Facility operated by the Oak Ridge National Laboratory under contract DE-AC05-00OR22725. 
This manuscript has been authored in part by UT-Battelle, LLC, under contract DE-AC05- 00OR22725 with the US Department of Energy (DOE). The publisher acknowledges the US government license to provide public access under the DOE Public Access Plan (http://energy.gov/downloads/doe-public-access-plan).

During the preparation of this work the author used OpenAI (GPT-5) in order to search literature and revise language. After using this tool/service, the author reviewed and edited the content as needed and takes full responsibility for the content of the published article.

\section*{Author contributions}
S.G. conceived the project, performed derivations, calculations, analysis, and wrote the manuscript.
\section*{Conflict of interest declaration}
The author declares no competing interests.

\bibliography{flexomag}

\end{document}